%% file: report-acm.tex
\documentclass[acmsmall]{acmart}

\AtBeginDocument{%
  }

\setcopyright{acmcopyright}
\copyrightyear{2025}
\acmYear{2025}
\acmDOI{XXXXXXX.XXXXXXX}

\input{preamble}
\usepackage{comment}

\usepackage{booktabs,array,threeparttable}
\newcolumntype{Y}{>{\centering\arraybackslash}p{0.085\textwidth}} 
\renewcommand{\arraystretch}{1.2}

\includecomment{JournalOnly}  
\includecomment{ConferenceOnly}  
\excludecomment{TulipStyle}
\usepackage{booktabs}
\AtBeginDocument{%
	\let\toprule\midrule
	\let\bottomrule\midrule
}

\begin{document}

\title{Digital Privacy Under Attack: Challenges and Enablers}


\author{Baobao Song}
\affiliation{%
  \institution{Faculty of Engineering and IT,  University of Technology Sydney}
  \city{Ultimo, NSW 2007}
  \country{Australia}}
\email{baobao.song@student.uts.edu.au}
\orcid{0000-0001-5630-5661}




\author{Shiva Raj Pokhrel}
\affiliation{%
  \institution{School of IT,  Deakin University}
  \city{Geelong, VIC 3216}
  \country{Australia}}
\email{shiva.pokhrel@deakin.edu.au}
\orcid{0000-0001-5819-765X}

 \author{Mengyue Deng}
\email{mydeng@tulip.academy}
\affiliation{%
  \institution{Business School,
Hunan University}
\city{Changsha 410082}
\country{China}
}

\author{Qiujun Lan}
\affiliation{%
  \institution{Business School,
Hunan University}
\city{Changsha 410082}
\country{China}
  }
\email{lanqiujun@hnu.edu.cn}
\orcid{0000-0001-7523-9487}

\author{Robin Doss}
\affiliation{%
  \institution{Centre for Cyber Security Research and Innovation, 
Deakin University}
  \city{Geelong, VIC 3216}
  \country{Australia}}
\email{robin.doss@deakin.edu.au}
\orcid{0000-0001-6143-6850}

\author{Tianqing Zhu}
\affiliation{%
  \institution{Faculty of Data Science, City University of Macau}
  \city{Macau}
  \country{China}}
\email{Tqzhu@cityu.edu.mo}
\orcid{0000-0003-3411-7947}


\author{Gang Li}
\authornote{Corresponding Author}
\email{gang.li@deakin.edu.au}
\affiliation{%
  \institution{Centre for Cyber Security Research and Innovation, 
Deakin University}
  \city{Geelong, VIC 3216}
  \country{Australia}}
\orcid{0000-0003-1583-641X}




\renewcommand{\shortauthors}{B. Song et al.}


\input{abstract}


%




\maketitle


\input{mainbody.tex}
\bibliography{tuliplab,PrivacyAttack}
\bibliographystyle{ACM-Reference-Format}


\end{document}

%% file: preamble.tex
\newcount\DraftStatus  
\ifnum\DraftStatus=1
	\usepackage[draft,colorinlistoftodos,color=orange!30]{todonotes}
\else
	\usepackage[disable,colorinlistoftodos,color=blue!30]{todonotes}
\fi 
\makeatletter
 \providecommand\@dotsep{5}
 \def\listtodoname{List of Todos}
 \def\listoftodos{\@starttoc{tdo}\listtodoname}
 \makeatother

\usepackage{color}

\usepackage{graphicx}
\usepackage{algorithm}
\usepackage[noend]{algpseudocode}  
\usepackage{breqn}
\usepackage{bbding}
\usepackage{subfigure}
\usepackage{multirow}
\usepackage{psfrag}
\usepackage{url}
\usepackage{hyperref}
\usepackage{siunitx}
\usepackage{cleveref}
\usepackage{booktabs}
\usepackage{rotating}
\usepackage{colortbl}
\usepackage{threeparttable}
\usepackage{paralist}
\usepackage{epstopdf}
\usepackage{nag}
\usepackage{microtype}
\usepackage{siunitx}
\usepackage{nicefrac}
\usepackage{lipsum}
\usepackage[english]{babel}
\usepackage[pangram]{blindtext}
\usepackage{tikz}
\usepackage{colortbl}
\usepackage{xcolor}
\usetikzlibrary{fit,positioning,arrows.meta,shapes,arrows}

%% file: abstract.tex
\begin{abstract}
We present a comprehensive analysis of privacy attacks and countermeasures in data-driven systems. We systematically categorize attacks targeting three domains: anonymous data (linkage and structural attacks), statistical aggregates (reconstruction and differential attacks), and privacy-preserving models (extraction, reconstruction, membership inference, and inversion attacks). For each category, we analyze attack methodologies, adversary capabilities, and vulnerability mechanisms. We further evaluate countermeasures including perturbation techniques, randomization methods, query auditing, and model-level defenses, examining their effectiveness and inherent privacy-utility trade-offs. Our analysis reveals that while differential privacy offers strong theoretical guarantees, it faces implementation challenges and potential vulnerabilities to emerging attacks. We identify critical research directions and provide researchers and practitioners with a structured framework for understanding privacy resilience in increasingly complex data ecosystems.

\end{abstract}

%% file: mainbody.tex

\section{Introduction}\label{sec-intro}

In recent years, 
the volume of sensitive personal data 
and the motivation of malicious actors have increased significantly \cite{pal2020should}. 
Such data drives personalised experiences 
and shapes new dimensions of privacy \cite{li2021developers}. 
For instance, 
location-based services integrate user data 
into intelligent applications 
to deliver on-demand solutions. 
However, 
these advancements have not been matched by 
adequate awareness of privacy preservation, 
a growing concern for users. 
As more organisations and individuals share their data, 
the risk of unintentionally revealing sensitive information rises, 
potentially compromising user privacy. 
Unintentional disclosure and diminished utility of private data 
have been frequently reported over the past decade.
According to a Statista report, 
over 422 million records were exposed globally 
due to data breaches in the third quarter of 2024. 
\footnote{\textit{\href{https://www.statista.com/statistics/1307426/number-of-data-breaches-worldwide/}{Nov 8, 2024, www.statista.com}}}
Additionally, 
IBM reported that the average total cost of a data breach in 2024 reached \$4.88 million, 
marking a 10\% increase compared to the previous year.\footnote{\textit{\href{https://table.media/wp-content/uploads/2024/07/30132828/Cost-of-a-Data-Breach-Report-2024.pdf}{July 1, 2024, www.ibm.com}}}

Incidents such as the famous \textit{Hugo Awards} 2015 attack~\cite{eppstein2015hugo} 
have already raised alarming privacy concerns. 
Furthermore, 
\citet{jain2021online} summarized that
on social networking sites like \texttt{Twitter}/\texttt{X} and \texttt{Facebook}, 
although users can hide their real identity using an alias, 
a third party can discover the true identity 
by linking leaked information from these sites.


As noted above, 
although private information is not explicitly identified, 
privacy could be violated by analyzing the data. 
Failure to fully guarantee privacy will 
discourage users from sharing their personal data with others, 
which will significantly hinder the expected development of data science, 
digital society, and innovation in several sectors,
including health and other public applications~\cite{khan2019data}. 

\begin{figure}[t]
    \centering
    \includegraphics[width = 0.95\linewidth]{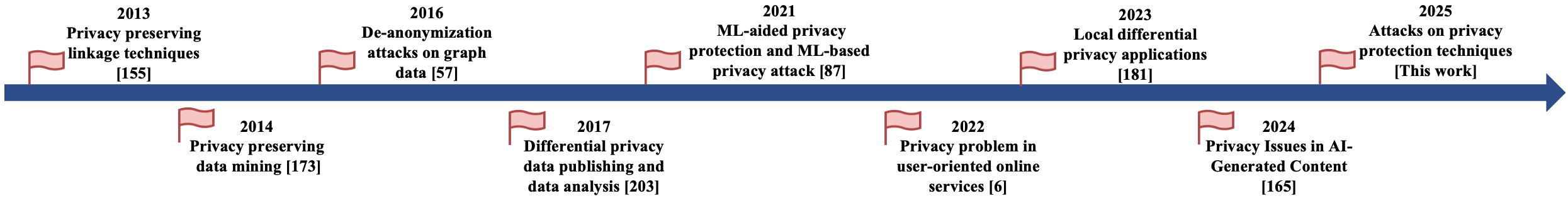}
    \caption{Progression of Privacy Preservation Surveys and Attack Strategies: 2013-2025} 
    \label{fig:timeline}
\end{figure} 

For a long time,
researchers have been trying to take a broader view of privacy issues 
and investigate different strategies  
to protect sensitive information from privacy attacks.
After 2012, many emerging issues 
have been studied extensively in the area of privacy protection, 
as illustrated in~\Cref{fig:timeline}, 
resulting in several valuable surveys, tutorials 
and important research articles~\cite{huang2023security, vatsalan2013taxonomy, leixu2014information, zhu2017differentially, ji2016graph, dwork2017exposed, yang2020local,liu2021machine,10.1145/3502288}. 
As shown in~\Cref{fig:timeline},
\Citet{vatsalan2013taxonomy} provided a detailed review 
of \emph{Privacy Preserving Linkage Techniques} (PPRL) 
which strike a balance between data utility and privacy.
\Citet{leixu2014information} reviewed the 
\emph{Privacy Preserving Data Mining}  technology 
at different stages of the \emph{Knowledge Discovery from the Data} process.
\Citet{zhu2017differentially} summarized the common issues 
related to data publishing and analysis in the context of \emph{Differential Privacy} (DP),
and~\citet{yang2020local} discussed 
\emph{Local Differential Privacy} (LDP) mechanism in different application scenarios~\cite{9141406}.

\emph{Machine learning} (ML) has been widely used 
in privacy research in recent years, 
as discussed and compared in detail in the recent survey~\cite{liu2021machine}. 
As a framework for distributed learning, 
\emph{Federated Learning} (FL) can protect local privacy without sharing it globally in the network.
\Citet{yin2021comprehensive} investigated privacy leakage risks in FL and 
introduced several privacy-preserving techniques. 
\citet{10.1145/3502288} reviewed the current approaches 
to privacy issues for online services 
through visualizations and design guidelines.
\citet{wang2024security}
examined the security and privacy challenges 
posed by AI-generated content in cyberspace, 
particularly the risks of privacy breaches and fabricated media.
In addition, 
cryptography is also an important tool for preserving privacy,
with advanced technologies such as 
\emph{General Secure Multi-party Computing} (SMPC), 
\emph{Privacy-preserving Set Operations},  
and \emph{Homomorphic Encryption} (HE). 
These technologies have been developed over a long period of time 
and have been systematically summarized 
by numerous privacy researchers~\cite{hasan2022privacy1}.
However, 
cryptography involves further encryption using mathematical approaches 
is beyond the scope of this paper. 
Of particular relevance to this study 
we now develop a high-level view of the notable milestones of several key aspects of \textit{personal privacy (other than encryption)}.

This paper addresses the protection of sensitive personal data in databases, a growing public concern~\cite{nissenbaum2020protecting}. Safeguarding such data is essential for individual autonomy amid rising risks from malicious inference and targeted attacks. While frameworks like FL and DP offer privacy-preserving solutions, their effectiveness varies across attack models. This study clarifies different privacy attack scenarios and evaluates existing tools, aiming to provide actionable insights for stronger data privacy defenses. Broader issues of organizational or national security are beyond the scope of this work.

Regarding previous surveys on privacy attacks, 
\citet{ji2016graph} discussed de-anonymisation attacks on graph data, 
their quantification, 
and impacts on data utility and privacy post-anonymisation. 
\citet{dwork2017exposed} examined privacy risks associated with aggregate data, 
focusing mainly on reconstruction and tracing attacks, 
including potential DP applications against such threats. 
\citet{liu2021machine} explored the intersection of machine learning (ML) 
and privacy, 
detailing ML-based privacy protection methods 
and ML-driven privacy attacks from varied perspectives.

Additionally, 
\citet{rigaki2020survey} categorised attacks targeting machine learning systems, 
while other studies concentrated specifically on 
particular attack methods or domains, 
including graph data de-anonymisation~\cite{ji2016graph}, 
social networks~\cite{shen2014defending}, 
blockchain~\cite{chen2022survey}, 
and location privacy attacks~\cite{gvili2020security}. 
Although existing surveys~\cite{ji2016graph, shen2014defending, chen2022survey, gvili2020security} 
have extensively covered specific privacy attack types, 
none has provided a holistic overview 
encompassing all privacy-related threats. Furthermore, 
some have evaluated the effectiveness of defensive strategies 
without adequately addressing the trade-off with 
data utility under privacy-preserving measures~\cite{salem2018ml, luo2023privacy}.


No recent comprehensive review is available
with a rigorous review of privacy preservation and attacks.  
In this work,
we address the gap in the literature
by providing a comprehensive survey on privacy attacks, 
along with mitigations and considerations of data utility
and discussions of impending challenges. 
Two key contributions are listed below.
\begin{enumerate}
\item 
We categorize existing research works 
on privacy attacks according to various criteria,
and identify detailed trends by 
comparing and analyzing the features of 
pioneering research 
for resilience to privacy attacks 
(see~\Cref{sec-preliminaries}).

\item 
We develop a systematic framework to explore and exploit the principles, 
methods,  preservation measures, and future research directions 
considering different types of privacy attacks.
We aim to provide a holistic understanding of the current development trends 
for the progress in data privacy research 
(see~\Cref{sec-anonyAttack} to~\Cref{sec-modelAttack}).


\end{enumerate}

The survey is organized as follows:
we provide some related concepts,
basic components and specific classification criteria
of privacy attacks in~\Cref{sec-preliminaries}.
In~\Cref{sec-anonyAttack}, 
we then summarize privacy attacks exploiting multiple techniques 
against anonymous data in detail.
We introduce existing privacy attacks targeted at statistical aggregate data publishing objects in~\Cref{sec-statisticalAttack}.
We discuss the privacy attacks embedded in various privacy-preserving models and mechanisms,
as well as interpreting the privacy risks contained in the privacy models
in~\Cref{sec-modelAttack}.
\Cref{sec-counter} presents the countermeasures 
for each of the mentioned attacks.
Moreover,
\Cref{sec-reandch} will envisage the promising emerging research issues and possible challenges related to the privacy attack.
Some concluding remarks will be provided in~\Cref{sec-conclusions}.

\section{Background and Survey Methodology} \label{sec-preliminaries}

Privacy encompasses a broad range of issues, 
including national security, business confidentiality, 
personal data protection, and individual autonomy. 
Fundamentally, privacy involves the right to safeguard personal information, 
thoughts, and activities from public scrutiny 
or unwanted interference~\cite{posner1977right}. 
In terms of national security, 
privacy safeguards state secrets and citizen safety; 
in business contexts, 
it prevents sensitive data exposure to competitors and cyber threats, 
preserving competitive advantage and customer trust. 
However, the most significant privacy implications affect individuals, 
emphasizing their right to control personal information—determining what is shared, 
with whom, and under what conditions~\cite{solove2012introduction}. 
This covers communications, financial records, 
health data, and location information.

\begin{figure}[t]
\centering
\includegraphics[width = 0.6\linewidth]{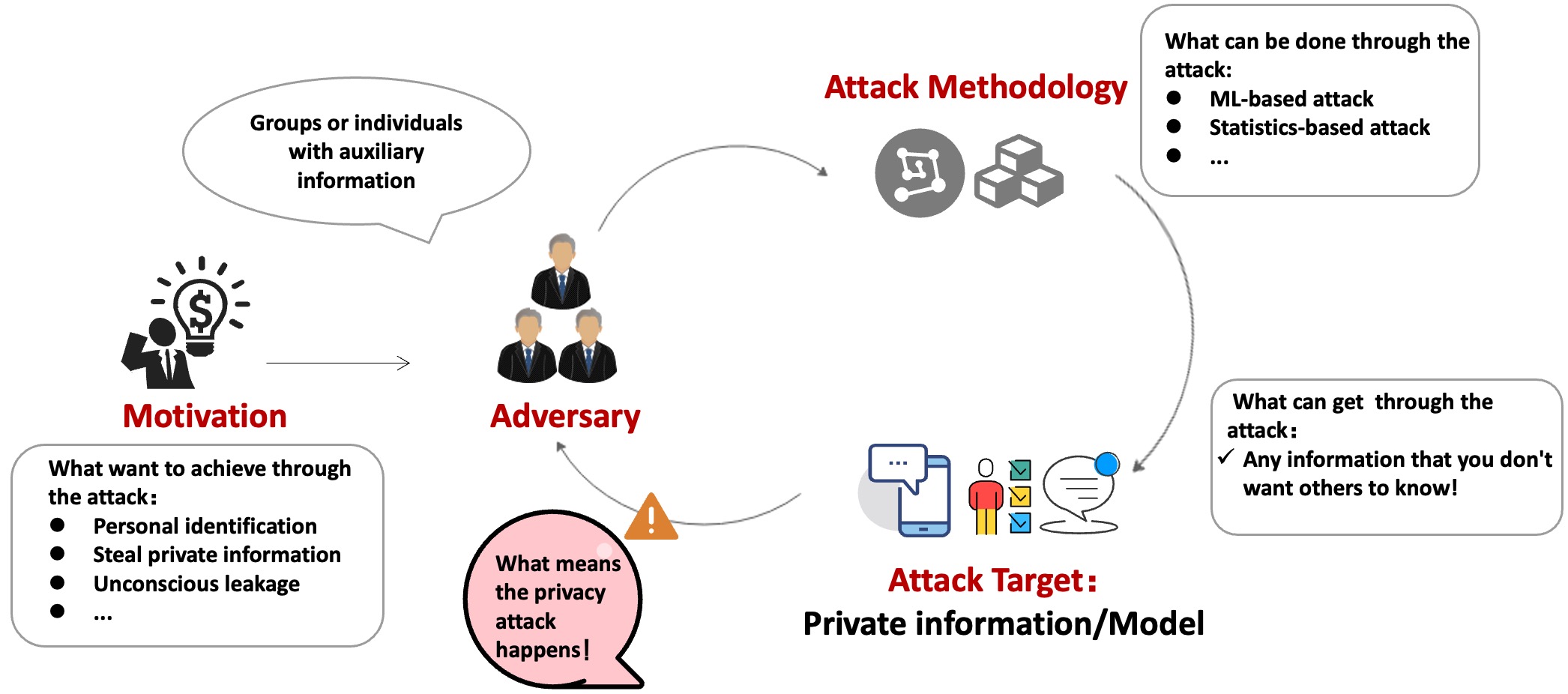}
\caption{The main components of the privacy attack process.}
\label{fig:attack}
\end{figure}

Recognising the paramount importance of individual privacy,
this paper specifically addresses protecting personal data within databases, 
highlighting privacy at the individual record level.
Before addressing privacy attacks, 
understanding privacy violations from an adversarial viewpoint 
is essential \cite{zhang2022privacy}. 
Adversaries breach privacy by acquiring information 
beyond what is publicly available, 
potentially causing harm. 
\citet{borisov2023application} describe examples such as linkage and reconstruction attacks, 
which identify private details by correlating anonymised data 
with known datasets or targeting hidden database information. 
Privacy violations occur when attackers exceed quantifiable thresholds, 
for example, by correctly matching 90\% of anonymised records. 
These attacks exploit seemingly harmless data releases 
to uncover sensitive information, 
as shown in \Cref{fig:attack}.

\subsection{Components of Privacy Attack}

\begin{figure}[htbp]
\centering
\includegraphics[width = 0.7\linewidth]{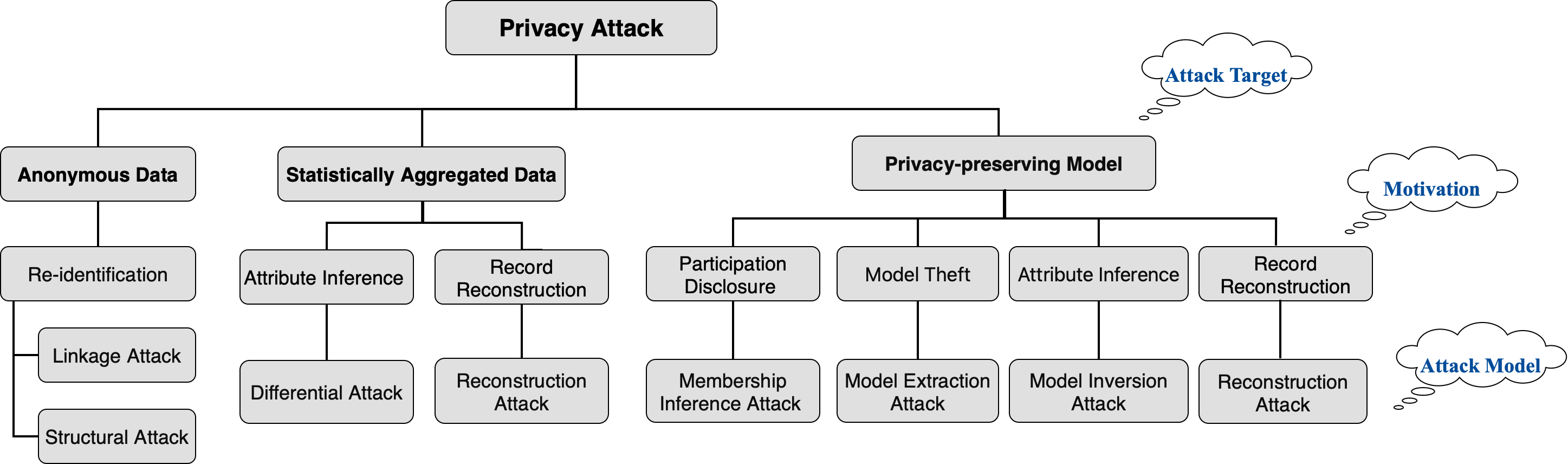}
\caption{High-level hierarchical view of privacy attacks}
 \label{fig:tree}
 \end{figure}

Four key components define the privacy attack process:
\begin{description}
\item[Adversary]
Adversaries—individuals or organizations 
conducting privacy attacks—use strategic methods 
involving auxiliary information such as background knowledge, 
query mechanisms, and publicly available data to breach privacy. 
The effectiveness of these attacks often depends on 
the adversary's auxiliary information from various sources. 
For example, linkage attacks have utilized public Facebook profiles 
and voter registration records \cite{minkus2015city}. 
However, some harmful attacks can occur 
even without such additional information \cite{pan2020privacy}.
\item[Motivation]
The likelihood and impact of privacy attacks 
significantly depend on the adversary's motivation, 
which is often challenging to quantify, 
especially in targeted attacks. 
For instance, 
celebrities' hospital visits might tempt healthcare staff to misuse patient records, 
causing privacy breaches. 
Such situations are increasingly common 
due to extensive data sharing and big data practices. 
Motivations may include personal gain, 
exposure of sensitive information, 
or unintentional privacy violations 
arising from broader data handling procedures.
\item[Target]
Targets of attacks include sensitive personal data 
and machine learning models, 
encompassing information individuals prefer to keep private, 
such as genetic data related to diseases \cite{all2024genomic}, 
user behavioral patterns \cite{di2021metaverse}, 
and sensitive personal relationships \cite{liu2021machine}. 
Privacy-preserving methods like data anonymization, 
statistical aggregation, data publishing, 
and encrypted machine learning models naturally attract attackers.
\item[Methodology]
Attack methodology is central to privacy breaches. 
Despite advancements in privacy-preserving techniques, 
new attack strategies continue to emerge, 
exploiting vulnerabilities in advanced algorithms \cite{lyu2022privacy,jin2022we}. 
This paper aims to comprehensively understand these attack methods 
and evaluate their resilience, 
guiding ongoing privacy research.
\end{description}
In this paper, 
we systematically summarize classic 
and recent privacy attacks using a multi-layered classification framework 
(\Cref{fig:tree}). 
Attacks are categorized by target components—anonymous data, 
statistically aggregated data, 
and privacy-preserving ML models—
and further differentiated by attack motivations, 
such as participant disclosure or attribute inference, 
as detailed in \Cref{tbl:classification}.

{\setlength{\heavyrulewidth}{0.4pt}
	\setlength{\lightrulewidth}{0.4pt}
	\setlength{\cmidrulewidth}{0.4pt}
	
\begin{table}[h] 
\tiny
  \centering
  \caption{Classification of Privacy Attacks}
  \label{tbl:classification}
\begin{tabular}{l|lll}
\toprule
Components   & \multicolumn{3}{c}{Classification Dimensions}   \\ 
\midrule
Adversary   & \multicolumn{3}{c}{\begin{tabular}[c]{@{}c@{}}Possessing Auxiliary Information, Lacking Auxiliary Information Information\end{tabular}}   \\
Target      & \multicolumn{3}{c}{\begin{tabular}[c]{@{}c@{}}Anonymous Data, Statistically Aggregated Data, Privacy-preserving ML Model\\\end{tabular}}   \\
Motivation  & \multicolumn{3}{c}{\begin{tabular}[c]{@{}c@{}}Re-identification, Participation Disclosure, Attribute Inference, etc.\\ 
\end{tabular}}   \\
Methodology & \multicolumn{3}{c}{\begin{tabular}[c]{@{}c@{}} Attacks Types (Reconstruction, Linkage, Differential, Structural, Model Extraction, etc.)\\
\end{tabular}}\\
\bottomrule
\end{tabular}
\end{table} 
}




\subsection{Privacy-Utility Trade-off}

\begin{figure}[b]
\centering
\includegraphics[width=0.55\linewidth]{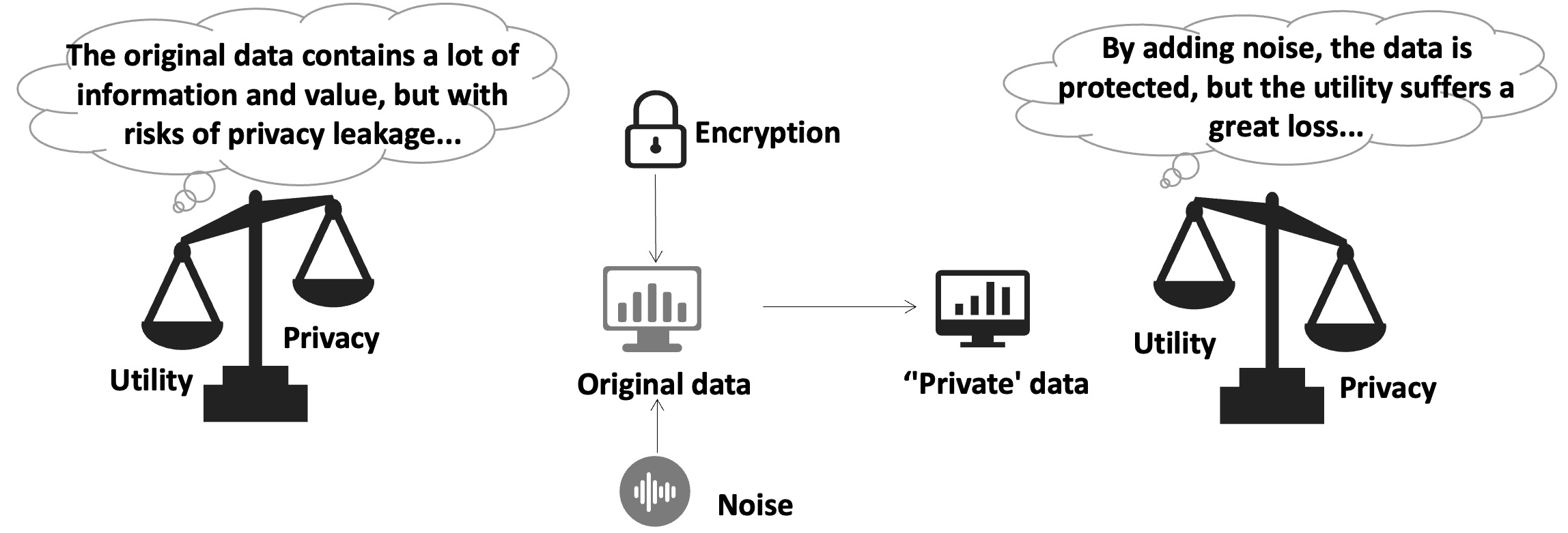}
\caption{Privacy-Utility Trade-off: 
increase in privacy gains decreases utility }
\label{fig:tradeoff}
\end{figure}

The trade-off between enhancing privacy and maintaining utility 
is widely recognized \cite{zamani2023privacy}.
Increasing the collection of personal data may increase commercial value,
but also increase the privacy risks. 
Ensuring privacy effectively, 
data curators often add noise to sensitive features, 
but this compromises utility and system performance,
known as the ``privacy-utility trade-off'' \cite{zamani2023privacy}. 
This trade-off is illustrated in \Cref{fig:tradeoff},
with the analytical value of the data being the main driver.

To this end,
\Citet{liao2019tunable} designed an adjustable measurement method 
for information leakage 
and explored its applications in privacy-utility trade-off.
\citet{zamani2023privacy} found the upper and lower bounds 
of the ``privacy-utility trade-off'' under 
the condition of satisfying bounded privacy constraints. 
The privacy-utility trade-off in noise-based privacy-preserving methods 
has been extensively explored and understood in the literature. 
Studies have focused on various differential privacy settings, 
aimed at balancing the privacy and utility implications 
over different types of private queries \cite{nanayakkara2022visualizing}. 
Further discussion on this aspect is provided in \Cref{sec-counter}.
%



\subsection{Survey Methodology: Knowledge Graph \& CiteSpace}

To conduct a thorough analysis of the literature, 
we employed a methodology that integrated 
Knowledge Graph~\cite{wang2017knowledge} 
and CiteSpace~\cite{zheng2023study} 
to analyze the current domain knowledge 
and research gaps in privacy attacks. 
CiteSpace can examine research connections 
and relationships among papers, 
and is used to create a knowledge graph~\cite{zheng2023study}. 
This effectively visualized complex relationships 
and trends within the academic literature of privacy attacks. 
We first analyzed and established exclusion criteria 
and filtered the downloaded literature. 
We then imported the final selection of publications into CiteSpace 
for cluster analysis to generate the knowledge graph.

\begin{figure}[!h]
\centering
\includegraphics[width =0.58\linewidth]{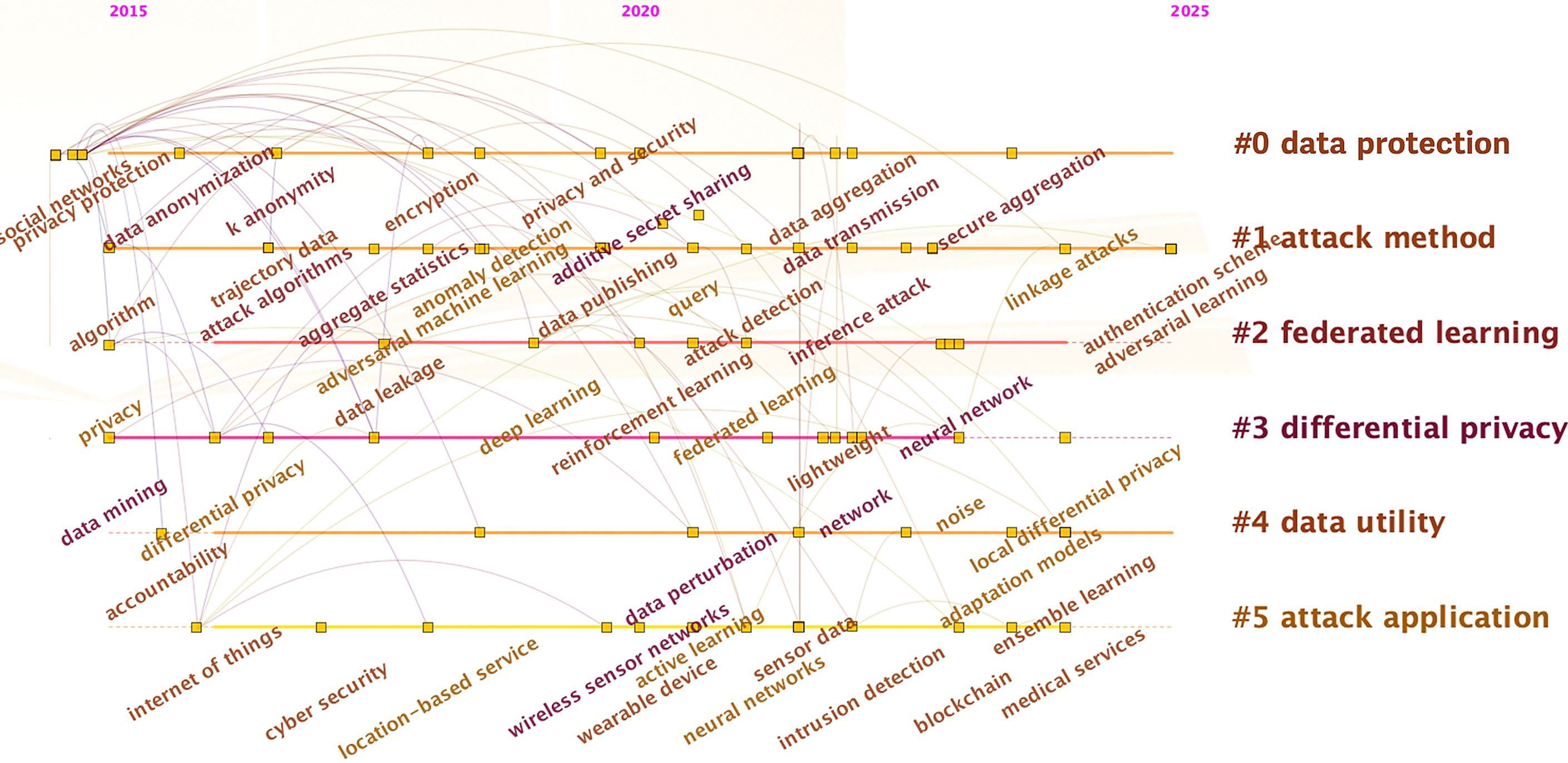}
\caption{Survey methodology with the Knowledge graph: timeline for privacy attack}
\label{fig:methedology}
\end{figure}

In the paper selection process, 
we initiated a search in the Web of Science Core Collection 
for publications related to the keyword ``personal data privacy attack'' 
over the last ten years, 
yielding \num{1450} papers. 
All non-English publications were excluded, 
leaving \num{1441} papers downloaded into the Zotero. 
After removing duplicates and retracted papers, 
we filtered the remaining papers, 
focusing on the specifically related attack topics. 
This narrowed the selection down to \num{1071} papers, 
screened on the basis of 
titles, abstracts, and keywords, 
excluding data papers and 
those not related to privacy attacks or 
did not address personal privacy data in databases.
Of the refined \num{1071} full-text papers assessed for eligibility, 
\num{25} were excluded due to inaccessible full text, 
irrelevant topics, different privacy protection objectives, 
or lack of experimental evaluations related to attacks, 
except for surveys, 
book chapters, and editorial materials. 
Finally, 
\num{1046} publications were included in the systematic review 
and imported into the CiteSpace.
In the CiteSpace, 
we performed clustering on the \num{1046} publications 
and used the timeline view to illustrate 
the knowledge graph (see ~\Cref{fig:methedology}). 
Seven clusters were generated, 
each representing key areas or popular contexts vulnerable to privacy attacks. 
We can see that Clusters \#0, \#1, \#2, 
and \#3 are related to attack methods and countermeasures, 
with Clusters \#2 and \#3 specifically representing 
widely recognized privacy-preserving approaches.
Meanwhile, Cluster \#4 focuses on the privacy-utility trade-off, 
while Cluster \#5 highlights areas of vulnerability.
This paper primarily examines various attack methods 
and defense strategies in various fields,
which our references have boiled down to \num{172}.

\section{Attacks on Anonymous Data} \label{sec-anonyAttack}

The conventional assumption in data privacy is that records containing sensitive information can be rendered anonymous by removing or modifying personally identifiable information (PII), such as names, Social Security numbers, or ID numbers. This process is referred to as de-identification, and the resulting records are commonly termed anonymous data. A widely adopted technique is the suppression of explicit identifiers (e.g., name, address, phone number), which are direct indicators of identity. However, extensive studies have demonstrated that this approach is insufficient, as individuals can often be re-identified by analyzing combinations of quasi-identifiers (QIDs)—attributes that are not inherently unique but, when linked together, can serve as strong proxies for identity. Examples of QIDs include demographic features (age, gender, ZIP code), behavioral traits (purchasing history, browsing activity), or medical characteristics. For instance, Sweeney’s seminal study showed that 87\% of the U.S. population could be uniquely identified using only three QIDs: ZIP code, gender, and date of birth. More recent analyses~\cite{rocher2021observatory,yan2023privacy} confirm that even sparse data attributes can, when combined, achieve near-unique matches, enabling re-identification attacks.

This phenomenon arises from the high dimensionality and sparsity of real-world datasets: as the number of attributes grows, the probability of an individual having a unique or rare combination increases significantly. Consequently, simply removing PII does not guarantee anonymity, since adversaries with auxiliary information (e.g., public voter rolls, social media data, leaked datasets) can link QIDs across databases to uncover identities.

Thus, while de-identification via PII removal is a foundational step, it is inadequate against modern re-identification risks. Recognizing the role of QIDs is central to understanding privacy vulnerabilities and motivates advanced protection techniques such as k-anonymity, l-diversity, t-closeness, differential privacy, and federated learning.

\begin{table}[!h]
\tiny
\caption{Anonymous Data}  
\centering  
\subtable[Original Data]{\label{tab:adataa}
\begin{tabular}{@{}llll@{}}
\toprule
Name  & \texttt{Age} & \texttt{Sex} & \texttt{Disease}  \\ \midrule
Alice & 26  & F   & \texttt{COVID-19}     \\
Bob   & 25  & M   & \texttt{COVID-19}     \\ 
Cathy & 26  & F   & \texttt{Diabetes} \\
David & 25  & M   & \texttt{Diabetes} \\
Eva   & 27  & M   & \texttt{AIDS}  \\ \bottomrule
\end{tabular}}
\qquad  
\subtable[Na\"ive Anonymized]{\label{tab:adatab}
     \begin{tabular}{@{}llll@{}}
\toprule
Id  & Age & Sex & Disease  \\ \midrule
1 & 26  & F   & \texttt{COVID-19}     \\
2   & 25  & M   & \texttt{COVID-19}     \\ 
3 & 26  & F   & \texttt{Diabetes} \\
4 & 25  & M   & \texttt{Diabetes} \\
5   & 27  & M   & \texttt{AIDS}   \\ \bottomrule
\end{tabular}}
 \label{tab:anonymousdata}  
\end{table}   

Let us consider a simple example,
where the information about the patients of a hospital 
is shown in~\Cref{tab:adataa}.
\Cref{tab:adatab} has been published by a curator, 
after deleting PIIs, 
in the hospital.
If the \textit{explicit identifiers} (i.e. name) are removed,
\textit{Alice} and \textit{Cathy} cannot be distinguished.
However,
if we know one of our neighbors,
a 27-year-old man,
has also visited the hospital,
we can conclude that he could be the one diagnosed with \texttt{AIDS}.
Here, 
(\textit{\texttt{Age}, \texttt{Sex}}) serve as QIDs.
The example above focuses on tabular data, 
but in reality, 
anonymous data spans a wide range of formats, 
each presenting distinct challenges. 
For instance, textual data, 
even after the removal of explicit identifiers, 
can inadvertently reveal identities through contextual information 
or distinctive writing styles that act as QIDs~\cite{manzanares2024evaluating}.
Similarly, 
visual data such as images or videos, 
despite the application of techniques like metadata stripping or face blurring, 
may still expose identities through subtle visual QIDs, 
such as unique objects, backgrounds, or physical characteristics~\cite{he2024diff}.
These examples illustrate the intricate complexities of maintaining privacy across diverse data types while ensuring the utility of the anonymized datasets.



\subsection{Linkage Attack}

Although each anonymized dataset individually may help 
to prevent the leakage of private information,
the combinations of several sanitized datasets,
known as \textit{linkage attack},
weaken such a guarantee.
In the linkage attacks on relational data,
the adversary (re)identifies individuals 
by linking the anonymized tables with some external tables 
that represent auxiliary information of individuals.
The linkage attack can be defined as follows.

\begin{definition}[Linkage Attack]\label{def:LA}
Let $D_1$ be a publicly available dataset 
containing quasi-identifiers $q$, 
and let $D_2$ be a private dataset containing both quasi-identifiers $q$ and sensitive attributes $s$.  
If a linkage function $f$ can map records from $D_1$ to $D_2$:  
$\exists~(r_1 \in D_1, r_2 \in D_2) \quad$ where $\quad r_1[q] = r_2[q]$,  
the adversary can successfully perform a linkage attack 
to infer the sensitive attribute $s$ for the individual associated with $r_2$.
\end{definition}


As a common privacy attack method exploited by malicious adversaries,
linkage attacks against anonymous data 
have been discussed in many research papers~\cite{sweeney1997weaving,sweeney2013matching,minkus2015city}.
As far as we know,
\citet{sweeney1997weaving} introduced the linkage attack 
for re-identification, firstly,
demonstrating the potential privacy risks of 
publishing Na\"ive anonymized data.
According to a famous statistical report on census data~\cite{sweeney2000simple},
researchers can uniquely identify 
\SI{87}{\percent} of the US population using the combinations of 
\textit{date of birth}, 
\textit{ZIP code}, 
and \textit{gender} with a high degree of probability and feasibility.
When databases do not share the full \emph{date of birth} with the public,
\citet{sweeney2013matching} further investigated the possibility
of matching known patients with anonymous health records
using \texttt{News} stories data.
Two common types of auxiliary information are outlined below.

\begin{description}
\item[Social networks]
The combination of personal information across multiple social networks
is usually referred to 
as \textit{online social footprint}~\cite{malhotra2012studying}.
Many researchers have focused efforts on 
social network matching using these combinations. 
In response to the challenge of users 
using personal information 
across different \emph{Online Social Networks} (OSNs), 
\citet{peled2016matching} have developed a sophisticated machine learning-based approach 
to accurately match user profiles across multiple platforms. 
In addition, 
automated identity theft attacks were addressed by \citet{bilge2009all}, 
which introduced a novel scoring system with a threshold 
to determine whether two specific accounts belong to the same user.
After that,
\citet{shen2014defending} presented the first countermeasure  
against user identity association attacks,
by developing a novel greedy algorithm  
to prevent identities on different OSNs from being linked.

\item[Trajectory]
Trajectory data is inherently high-dimensional, sparse, and sequential, typically represented as ordered spatio-temporal doublets. While high-quality trajectory data is vital for accurate data mining and predictive modeling, it simultaneously exposes highly sensitive details about individuals’ lifestyles, routines, and movements. This creates severe privacy risks: adversaries equipped with even partial trajectory information as auxiliary knowledge can reliably compromise anonymity.

The absolute uniqueness of human mobility was rigorously demonstrated by~\citet{demontjoye2013unique}, who derived a formal expression linking data resolution and the number of known spatio-temporal points to quantify re-identification risk and establish privacy bounds. Extending this, \citet{xu2019no} showed that even seemingly benign mobility traces, such as social media check-ins, can enable high-probability re-identification when correlated with other mobility datasets. Together, these findings confirm that human mobility patterns are intrinsically identifiable, making trajectory privacy one of the most challenging aspects of data protection.

\end{description}

In addition to the above, 
we should also be aware that auxiliary information may, 
in many circumstances, 
pose a privacy risk to some extent due to its uniqueness and distinctiveness,
especially when combined with other data.
For example, in social link mining, 
check-in information is often used as auxiliary data 
to reveal potential social relationships among travelers~\cite{vu2019breach}. 
Similarly, in data mining, 
shopping receipt data frequently serves as auxiliary knowledge 
to construct detailed customer profiles~\cite{gursoy2021recovering}.
In these cases, 
both check-in information and shopping receipt data 
inherently contain privacy-sensitive personal details. 
When analyzed in conjunction with other datasets, 
they can further amplify privacy risks, 
leading to the exposure of even more sensitive information.

Early privacy attack research often assumed auxiliary information was confined to predefined QIDs. However, \citet{narayanan2007how} demonstrated powerful de-anonymization attacks on high-dimensional micro-data (e.g., transactions, movie ratings), effective even on sanitized datasets and tolerant to noisy auxiliary knowledge. Building on \citet{merener2012theoretical}, they showed that rare attributes amplify re-identification power, cutting the required auxiliary information by up to 50\%. The discontinuation of the Netflix Prize~\cite{singel2010netflix} highlighted the severity of such risks, proving that simple anonymization cannot ensure privacy.

In today’s big data environment, linkage attacks pose even greater threats. \citet{zheng2018data} showed how data from different devices, dimensions, and participants in IoT systems can be cross-linked to compromise privacy. In digital finance, \citet{zhang2020deanonymization} revealed that combining physical and virtual payment flows enables precise transaction re-identification. Similarly, \citet{christen2018pattern} proposed a pattern-mining-based linkage attack on encrypted records, later refined by \citet{nobrega2022explanation} to overcome BC-PPRL limitations. In emerging environments such as the vehicular metaverse, \citet{luo2023privacy} warned that synchronization and large-scale communication among Vehicular Twins could expose both identity and location. \Cref{tbl:linkageSource} summarizes the auxiliary information exploited in well-known linkage attacks.

\begin{table}[t]
\tiny
\arrayrulecolor{black} 
  \centering
  \caption{Existing works and dimension of linkage attacks}
  \label{tbl:linkageSource}
  \resizebox{\textwidth}{!}{
  \begin{tabular}{p{100pt}<{\centering}|p{100pt}<{\centering}p{100pt}<{\centering}p{100pt}<{\centering}}
\toprule
    References & Data Type & \multicolumn{2}{c}{Data Resources} \\
\midrule

\citet{sweeney1997weaving}\cite{sweeney2000simple} & Tabular data & Anonymous medical data & Public census data \\
   \citet{zang2011anonymization} &Trajectory data& Anonymous location data & Public census data \\
   \citet{minkus2015city} & Tabular data and Graph data & Public Facebook profiles & Voter registration records \\
   \citet{xu2019no} & Trajectory data& Check-in records & Additional mobility data \\
   \citet{zheng2018data} & Hybrid data & Contents from different dimensions & Contents from different participants \\
   \citet{gursoy2021recovering} & Tabular data and Sequence data & Sex or ancestry & Allele-specific genes \\
    \citet{christen2018pattern} &  Encoded data & Bloom filter encoding  & Sets
of Bloom ﬁlters \\
\citet{srivastava2020evaluating} &  Audio data & Anonymized speech recordings & Enrollment set and Public speech datasets\\

 \bottomrule
\end{tabular}}
\end{table}

\subsection{Structural Attack}

Graph data is widely shared in areas like mobility traces and social networks, 
yet privacy concerns require anonymization—often at the expense of data utility.
Improper anonymization of graph data 
will lead to the degradation of data utility, 
because it greatly reduces the information values 
contained in nodes and edges during graph analysis~\cite{li2023private}. 
The attack methods aim to re-identity 
the anonymized users from the anonymized graph, 
so-called \textit{structural attack}~\cite{zhao2021structural}.

\begin{definition}[Structural Attack]\label{def:SA}

Let $G= (V, E)$ be a graph, 
where $V$ is the node set (representing individuals) 
and $E\subseteq V\times V$ is the edge set corresponding to
the relationships between different nodes.
The goal of the structural attack is to construct a mapping function $f$ 
to re-identify or infer sensitive attributes of individuals.

\end{definition}

Graph de-anonymization reconstructs or partially 
recovers an original graph from its anonymized version 
by matching node or edge properties. 
Recent structural de-anonymization methods \cite{zhao2021structural, li2020privacy} 
re-identify anonymized subjects using unique structural features 
and external auxiliary knowledge. 
These approaches can be grouped broadly into structural information-based 
and seed-based methods. 
While structural methods operate autonomously, 
seed-based techniques rely on identified users (``seeds'') to guide the attack. 
Furthermore, dynamic graphs—with continuously 
evolving structures—remain vulnerable to these attacks, 
especially when adversaries combine structural strategies 
with models like game theory to optimize interactions between attackers 
and defenders~\cite{chen2020active}.


\subsubsection{Structural information-based de-anonymization}

Re-identifying anonymized nodes often hinges on auxiliary information. \citet{zhou2008brief} distinguish six key data types for privacy attacks: 
node attributes, node degrees, link relationships, 
individual neighborhoods, embedded subgraphs, and graph metrics. 
\citet{hay2008resisting} further specify three adversarial knowledge categories—node refinement queries, 
subgraph queries, and hub fingerprint queries—that can undermine naive anonymization.

Using knowledge graphs to bolster auxiliary information, 
\citet{qian2019social} demonstrate enhanced de-anonymization 
and inference attacks in social networks. 
Similarly, \citet{zhang2019attributeenhanced} emphasize the role of user attributes 
in enabling more potent attacks 
by constructing multipartite graphs and mapping networks. 
\citet{fu2023TEMPeffect} show that increasing network symmetry 
reduces de-anonymization capability 
by introducing greater uncertainty in node matching. 
Beyond structural attacks alone, the 
proliferation of big data suggests that 
malicious exploitation of additional auxiliary information 
can lead to even more powerful privacy attacks—an area 
demanding further investigation.

\subsubsection{Seed-based de-anonymization}

\begin{figure}[ht]
\centering
\includegraphics[width = 0.37\linewidth]{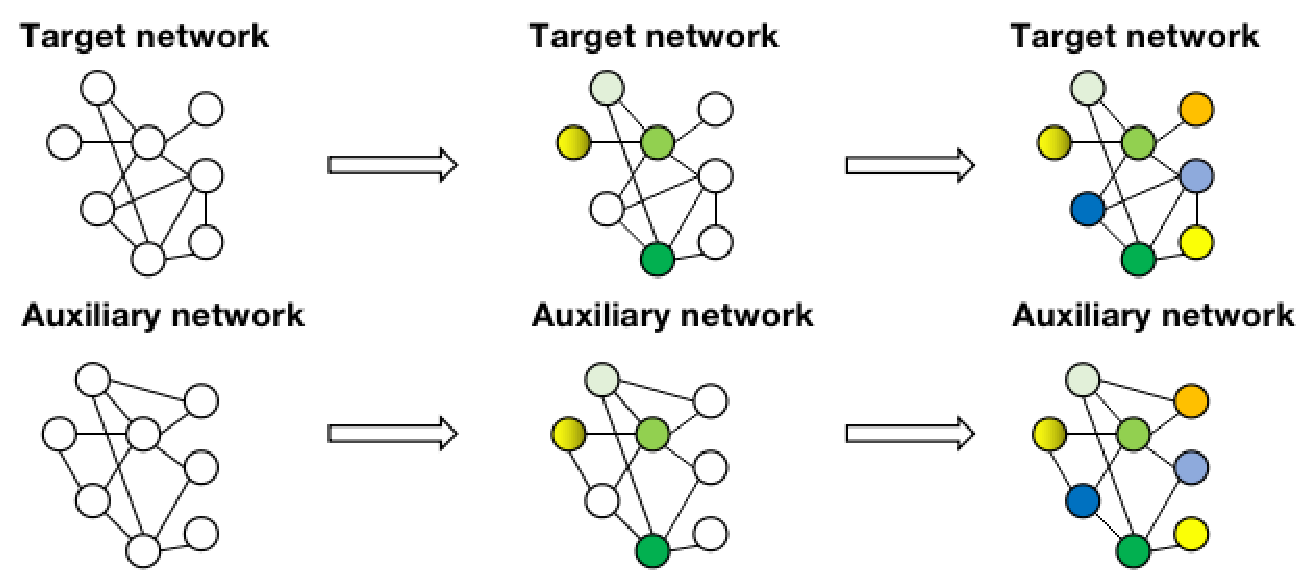}
\caption{An example of seed-based de-anonymization where
1) \textit{Seed identification}: mapping some seeds between two networks 
through unique subgraph pattern search; 
2) \textit{Propagation}:
expanding the set of matched users by comparing 
and mapping the neighbors of previously matched seeds incrementally. }
\label{fig:seed}
\end{figure}

One of the first works to study structural information-based 
de-anonymization attacks discussed active and passive attacks using small subgraphs 
designed to violate the privacy of social network users~\cite{backstrom2007wherefore}.
While providing an important reference,
due to the limitations of practicality and effectiveness,
\citet{narayanan2009deanonymizing} introduced a classic and widely applied approach to de-anonymization,
which models the process of de-anonymization as two steps:
\textit{seed identification} and \textit{propagation}. 
The seed represents a node in a graph 
that can provide some individual auxiliary information. 
The method performs a `network alignment',
which matches the anonymized graph's nodes
with the nodes of the auxiliary graph 
that have known identities as accurately as possible.
\Cref{fig:seed} illustrates the process of seed-based de-anonymization.

More importantly,
motivated by this work,
a series of seed-based de-anonymization attacks emerged~\cite{yartseva2013performance,weipeng2014twostage}.
\citet{yartseva2013performance} further improved the attack 
in~\cite{narayanan2009deanonymizing},
and presented a simple \textit{percolation-based} 
de-anonymization algorithm.
It attempted to match each pair of nodes
from the auxiliary and anonymized graphs,
where both nodes have at least $r$ neighbor pairs. 
\citet{weipeng2014twostage} introduced 
a two-stage algorithm called \textit{Seed-and-Grow}
to achieve the re-identification of anonymous individuals within social networks,
two stages of which were based on the graph structure and
overlapping user bases between different social networks respectively. 
To improve accuracy and efficiency, 
\citet{zhang2021efficient} introduced a new framework
for node matching by quantifying the matching score of a pair, 
using higher-order neighboring information.
However, 
these seed-based de-anonymization techniques require a large number of seeds, 
and they have difficulty effectively mapping the nodes of the large-scale anonymized graph.


To overcome these shortcomings,
\citet{nilizadeh2014communityenhanced} proposed a 
\textit{community-enhanced} de-anonymization attack
with the nature of ``\emph{divide-and-conquer'}'.
It performs the mapping at the community level for the first time,
after which the heuristic 
network mapping method~\cite{narayanan2009deanonymizing}
is then applied to the nodes within the de-anonymized communities,
and finally to the whole graph.
A Bayesian attack framework focusing on 
seed-free graph de-anonymization 
has been considered by~\citet{pedarsani2013bayesian},
which does not require any side information or initially mapped nodes.
\citet{shao2019fast} proposed 
a seedless de-anonymization method called \textit{RoleMatch}, 
which uses node similarity and neighborhood matching 
to achieve efficient node matching.
A practical \emph{Unified Similarity} (US) measurement
used in the mapping propagation step 
has been defined by~\citet{ji2016general},
and a US-based de-anonymization framework 
has been generalized to an adaptive framework,
which eliminates the need for the adversary 
to obtain auxiliary information about the amount of overlap 
between the auxiliary dataset and the anonymous dataset. 
\Cref{tbl:structural} compares some representative works on structural attacks
with their contributions and limitations.

\begin{table}[t] 
\arrayrulecolor{black}
  \centering
  \caption{Comparison of the Methods of Structural Attacks and Limitations}
  \label{tbl:structural}
\resizebox{\textwidth}{!}{
 \begin{tabular}
 {m{100pt}<{\centering}|m{180pt}<{\centering}m{145pt}<{\centering}m{130pt}<{\centering}}
\toprule
Representative work & Method & Contribution & Limitation \\ \midrule
\citet{zhu2017differentially} & Neighborhood attacks 
& Adversary exploits the background knowledge about neighbors 
& Can be defended by DP \\ \midrule 
\citet{backstrom2007wherefore} 
& Design active attacks (disturb the links to neighbors) and passive attacks (de-anonymize the neighbors) to graph data 
& Adopt neighboring information to identify matching pairs effectively 
& Not scalable and can be defensed \\ \midrule
\citet{yartseva2013performance} 
& Percolation graph matching 
& Adopt neighboring information to identify matching pairs effectively 
& Wrong matching by only relying on local information \\ \midrule
\citet{zhang2021efficient} & Personalized PageRank-based Graph Matching  
& Apply Personalized PageRank to quantify the pair of nodes' matching score 
& The dependence of seed \\ \midrule
\citet{pedarsani2013bayesian} 
& Bayesian method for approximate graph matching 
& Seed-free attack 
& Performance decreases significantly as the graph density increases \\ \midrule
\textcolor{black}{\citet{chen2020active}} 
& \textcolor{black}{Dynamic re-identification attack for exploiting tempo-structural patterns} 
& \textcolor{black}{A game-theoretic approach to optimize attack strategies in dynamic graphs} 
& \textcolor{black}{Rely on sybil patterns} \\
\bottomrule
\end{tabular}}
\end{table}

\section{Attacks on Statistical Aggregated Data} \label{sec-statisticalAttack}

Nowadays,
due to the sheer volume of data collected,
many databases contain confidential information.
The data curator may be reluctant to release the original data directly 
or share it completely with untrusted parties 
for privacy reasons. 
For example, 
data curators may only allow statistical and aggregate queries 
to these sensitive databases,
or more cautiously,
choose to publish noisy or encrypted databases~\cite{naveed2015inference}.
In recent years,
extensive research has shown that 
the adversary who monitors various query results of the databases
can obtain private information,
including inferring the sensitive values of data
or even reconstruct them~\cite{kellaris2016generic}. 
Maintaining confidentiality 
by publishing statistical aggregates,
while protecting privacy to some extent
is usually very challenging. 
This is because various statistics or query results
relate the sensitive information to potentially 
with public information,
which may inadvertently reveal private information,
resulting in privacy breaches.
It is important to note that 
statistical data is not limited to tabular data alone. 
Other forms, such as time series data~\cite{voyez2022membership} and trajectory data~\cite{tu2018new}
also fall under the category of statistical data. 
These data types share the common characteristic 
of being analyzable through statistical methods, 
yet each presents different structural complexities 
and challenges in terms of privacy preservation.
These observations raise critical questions:
\textit{Can aggregate statistics be published without compromising individual privacy? Is it possible to infer sensitive information solely from such aggregates? Under what conditions can adversaries reconstruct most of a database from aggregate query outputs? Does executing encrypted queries on fully encrypted databases ensure true privacy?}

Motivated by these questions, recent research has focused on attacks against aggregated datasets, including reconstruction attacks\cite{dwork2017exposed,kellaris2016generic}, differential attacks\cite{zhao2020novel}, and membership inference attacks~\cite{pyrgelis2017knock}, each revealing fundamental vulnerabilities in statistical data publication.
\color{black}

\subsection{Reconstruction Attacks}
\label{recAttack}

\Citet{dinur2003revealing} first reported the reconstruction attack, 
which attempts to reconstruct the statistical database 
from the private query results of linear statistics.
After that,
a large number of follow-up works investigated the reconstruction attack~\cite{dwork2017exposed},
which provides favorable guidance in the theoretical development of 
a rigorous approach to the private publishing of statistical aggregate data. 
With the continuous advances and research interests in the field, 
reconstruction attacks on various types of original data,
such as social links~\cite{ghasemikomishani2016pptd},
encrypted data~\cite{kellaris2016generic} 
or image data~\cite{mai2019reconstruction},
are widespread. 
The reconstruction attack can be defined as follows~\cite{dinur2003revealing,dwork2017exposed}. 

\begin{definition}[Reconstruction Attack]\label{def:RA}
Consider a database with $n$-record,
where each record contains a unique identifier and information ($q_{i}$, $s_{i}$).
The sensitive bit, denoted by $s_{i}$,
contains some private information,
while the remaining parts of the record,
denoted by $q_{i}$,
can be considered as available and public.
Given that $s$ is the column vector of 
the sensitive bits in the database,
the goal of the reconstruction attack is to
generate a vector $s'$ of all $n$ bits 
that match $s$ as closely as possible.
\end{definition}

From~\Cref{def:RA},
we can conclude that 
reconstruction attack attempts to determine the sensitive bits 
of individuals in the database 
on which the private queries are performed.
By using multiple query methods 
to infer the sensitive value, 
reconstruction attack methods can be broadly divided into 
linear query and range query,  
as discussed later in \Cref{recAttackLQ} and \Cref{reconstructionrq} respectively. 
\textcolor{black}{
In addition to the two mainstream methods, 
there are also reconstruction attacks specifically targeting privacy-preserving ML models, 
which are discussed in ~\Cref{mlrecon}.}


\subsubsection{Linear Query for Reconstruction Attacks}
\label{recAttackLQ}

By executing noisy statistical queries on the database of $n$ records
together with using the results to 
create a set of mathematical constraints,
the reconstruction attacks can be transformed into 
the task of solving simultaneous equations.
That is,
there usually exists a linear programme that can reconstruct 
most of the fraction of database~\cite{garfinkel2019understanding}. 
In their exploration, 
\citet{dwork2017exposed} delved into the approximation of a secret vector $s$
using publicly available statistics $\hat{q}$ as an approximation for the matrix $B$. 
Each record of $B$ corresponds to specific queries. 
Thus, 
reconstruction attacks based on linear statistics 
revolves around understanding the $B$-reconstruction problem. 
This involves deciphering whether the secret vector $s$ 
can be derived from $\hat{q}$, 
given a collection of statistical data represented by the records of the matrix $B$.



Many studies are devoted to 
achieving the trade-off between utility and privacy 
in linear query for reconstruction attacks~\cite{dinur2003revealing, dwork2008new}.
The seminal study of~\citet{dinur2003revealing}
introduced a polynomial-time algorithm $M$  to
reconstruct a good approximation to the statistical database
with low error and a high success probability. 
They drew a pivotal conclusion:
to prevent \textit{blatant non-privacy} as defined above, 
it is necessary to add the perturbation of magnitude 
$\Omega (\sqrt{n})$ to the output perturbation sanitiser,
which may completely break the database utility.
~\Cref{tbl:recfirst} summarizes the main results of this pioneering work.
In summary,
\citet{dinur2003revealing} demonstrated the trade-off 
between data utility and privacy 
when computing statistics on sensitive information 
from the confidential database.


{\setlength{\heavyrulewidth}{0.4pt}
	\setlength{\lightrulewidth}{0.4pt}
	\setlength{\cmidrulewidth}{0.4pt}

\begin{table}[h]
  \centering 
  \fontsize{6}{6}\selectfont
   \resizebox{\textwidth}{!}{
  \begin{threeparttable}
  \captionsetup{font=scriptsize, labelfont=scriptsize}
  \caption{\scriptsize Perturbation and Privacy}
  \label{tbl:recfirst}
  \begin{tabular}{m{90pt}<{\centering}|m{90pt}<{\centering}m{100pt}<{\centering}}
\toprule
    Perturbation of query algorithm $A$ & Privacy of statistical database $D$ & Adversary \\
\midrule
   $o(n)$ &  $\exp(n)$-non-private & Exponential Adversary \\
   $o(\sqrt{n})$ & poly(n)-non-private & Polynomially Bounded Adversary \\
   $o(\sqrt{T(n)})$ & ($T(n)$,$\delta$)-private\tnote{1} & Time-$T$ Bounded Adversary \\
 \bottomrule
\end{tabular}
 \begin{tablenotes}
        \item[1] $T(n)>polylog(n)$,$\delta>0$.
      \end{tablenotes}
  \end{threeparttable}
  }
 \end{table} 
}
After that,
a number of studies were carried out on 
the effects of the perturbation added to the answers of the queries,
the limitations of adversary queries
and the computational time complexity
on the success of the reconstruction attacks
~\cite{dwork2007price,dwork2008new}. 
\citet{dwork2007price} introduced a somewhat extreme situation 
where the curators give completely wrong answers 
to a small fraction of queries,
by combining the reconstruction attacks with 
\textit{LP Decoding} for error correction.
They showed that 
any database query mechanism that
provides arbitrarily inaccurate answers on a \num{0.239} fraction 
of randomly generated privacy weighted subset-sum queries and adds additional $O(\sqrt{n})$ error 
to any reasonable number of answers
can be non-private.
\citet{dwork2008new} introduced a class of more powerful attacks
that require only $n$ deterministically chosen queries.
For these queries,
adding an arbitrary perturbation to 
a $(\frac{1}{2}-\varepsilon)$ fraction of the responses 
can not defend against an adversary 
who completes the attack in a fixed amount of time, 
regardless of the size of the database.
The \textit{No-free-lunch} theorem~\cite{kifer2011no}
indicates that the published responses to statistical queries 
provide evidence of data participation,
so that 
the subsequent noise queries will not be sufficient to protect privacy.
From above,
we should know that the accuracy of reconstruction attacks
depends on the magnitude of the perturbation,
the number of queries and the size of the database.
\Cref{tbl:reccompare} summarized the main features of the above 
representative reconstruction attacks.

{\setlength{\heavyrulewidth}{0.4pt}%
	\setlength{\lightrulewidth}{0.4pt}%
	\setlength{\cmidrulewidth}{0.4pt}%
	\setlength{\arrayrulewidth}{0.4pt}%
	
\begin{table}
  \centering 
  \fontsize{6}{6}\selectfont
  \caption{Main Features of Representative Reconstruction Attacks}
  \label{tbl:reccompare}
  \resizebox{\textwidth}{!}{
  \begin{tabular}{m{70pt}<{\centering}|m{40pt}<{\centering}m{45pt}<{\centering}m{100pt}<{\centering}}
\toprule \tiny
    References & Query & Runtime & Noise\\
\midrule
   \Citet{dinur2003revealing} & $O(n{\log ^2}n)$ & $O(n^5\log^4{n})$ & $O(\sqrt{n})$\\
   \Citet{dwork2007price} & $O(n)$ & $O(n^5)$ & 0.239: arbitrarily inaccurate \\
   \Citet{dwork2008new} & $O(n)$ &  $O(n\log{n})$ & $O(\sqrt{n})$ \\
   \Citet{dwork2008new} & $O(n\log{n})$ & $Poly(\frac{e}{\varepsilon })$ & $\frac{1}{2}-\epsilon$: arbitrarily inaccurate\\
 \bottomrule
\end{tabular}}
\end{table}
}

To further confirm the practical feasibility and effectiveness 
of common reconstruction attacks 
and understand the resilience of statistical aggregate publishing 
to these attacks,
numerous research works have investigated the possibility of 
bringing reconstruction attacks into practice~\cite{garfinkel2019understanding, cohen2018linear}.
For example, 
the protections for the 2010 U.S. Census data failed, 
resulting in the leakage of many individuals' information.
\Citet{abowd20232010} proved that by conducting reconstruction attacks, 
attackers are able to infer the responses of 
\num{3.4} million vulnerable individuals with 95\% accuracy.
The case of databases containing  
health statistical data 
were discussed by~\citet{vaidya2013identifying}.
They proposed
a possible identifying inference attack via \textit{HCUPnet},
a free,
online,
privacy query system based on health statistical data from \textit{HUCP}.
They combined the results of privacy statistical queries with 
integer programming technology,
and then successfully realized the reconstruction 
attacks against healthcare databases.
With the continuous updating and optimization of 
some \textit{NP-hard} solvers and the great improvement in computer speed,
it is not just a theoretical risk to carry out reconstruction attacks 
on larger and more complex databases.


\subsubsection{Range Query for Reconstruction Attacks}\label{reconstructionrq}
Beyond aggregate data releases, a common strategy is to outsource massive datasets to third-party servers. Cryptographic techniques enable confidentiality while permitting efficient encrypted queries and responses~\cite{yang2018expressive}. However, achieving a practical balance between privacy and efficiency remains challenging~\cite{gui2019encrypted}. Recent studies further reveal that confidentiality can be compromised when adversaries gain access to auxiliary information, such as large volumes of encrypted queries and results~\cite{li2021privacy1}.
\color{black}

\begin{table} [h] 
\tiny
  \centering
  \caption{Comparisons of Reconstruction Attacks}
  \label{tbl:rec-compare}
  \resizebox{\textwidth}{!}{
  	{\setlength{\heavyrulewidth}{0.4pt}
  		\setlength{\lightrulewidth}{0.4pt}
  		\setlength{\cmidrulewidth}{0.4pt}
  \begin{tabular}{p{40pt}<{\centering}|p{155pt}<{\centering}p{155pt}<{\centering}}
\toprule
     Aspect & Reconstruction Attacks based On Linear Querying 
     & Reconstruction Attacks based On Range Querying\\
\midrule
   Adversary & Honest but Curious Adversary & Persistent Passive Adversary \\
   Motivation & Reconstruct the Sensitive Bits & Crack Encrypted Information \\
   Target &  Perturbed Statistical Database & Encrypted Database \\
   Methods & Exploit query results of linear statistics & Exploit special results of range querying\\
 \bottomrule
\end{tabular}}}
\end{table} 

Unlike reconstruction attacks based on linear queries 
as discussed in~\Cref{recAttackLQ},
only if a significant fraction of the encrypted data
can be accurately reconstructed with high probability in polynomial time, 
it can be claimed that 
the reconstruction attacks based on range queries are successful. 
\Cref{tbl:rec-compare} compares and summarizes 
these two types of reconstruction attacks.
The recent series of range query based reconstruction attacks
have greatly increased the risks to the privacy of 
personal data stored in encrypted databases~\cite{markatou2023attacks,wang2023fpmrq}. 

\begin{definition}[Range Query]\label{def:RQ}
A \textit{range query} $[a, b]$ initiated on an encrypted database 
will return an \textit{identifier} set
$M=\left \{r \in R: var(r) \in [a,b]\right \}$,
where  both $a$ and $b$ are integers, 
and $a \leq b$.
\end{definition}



Databases supporting range queries are especially vulnerable, as query results can inadvertently expose sensitive details about the encrypted data. As defined in \Cref{def:RQ}, we first consider volume attacks, where the adversary observes only the number of records returned. Such attacks effectively reconstruct database counts: while they do not map individual records to values, the recovered counts alone can cause severe privacy leakage. For instance, exact counts may reveal the distribution of credit ratings in a bank or salary levels of key employees. Publishing precise record counts thus creates subtle yet critical privacy risks, underscoring the need for advancing modern privacy-preserving techniques.

The work of \citet{kellaris2016generic} was the first systematic study of volume attacks and full database reconstruction from range queries. However, their approach relied on strong uniformity assumptions and required an impractically large number of queries, making it largely conceptual. Building on this, \citet{grubbs2018pump} simplified reconstruction to a clique search problem, operating under weaker assumptions. \citet{gui2019encrypted} further advanced the field with a robust attack resilient to fake requests, fake responses, and injected noise. More recently, \citet{poddar2020practical} demonstrated highly practical volume-based attacks, capable of compromising Gmail within minutes, and proposed three auxiliary strategies to bypass existing mitigations.

Collectively, these works show that volume leakage poses serious real-world threats, far beyond theoretical interest. Hiding access patterns alone is insufficient—developing stronger, principled privacy-preserving countermeasures is essential.
\color{black}

In addition to volume reconstruction attacks that exploit volume leakage, 
many other information leaks can be exploited to achieve 
encrypted database reconstruction 
under different query conditions.
\citet{lacharite2018improved} discussed the conditions 
under which the adversary's auxiliary information 
is bounded in \textit{access pattern leakage}
and \textit{rank information leakage}.
Within this auxiliary information,
three types of reconstruction attacks have been presented.
\Citet{grubbs2019learning}'s work combined the ideology
of statistical learning theory,
introducing $\varepsilon$-approximate 
database reconstruction ($\varepsilon-ADR$)
and $\varepsilon$-approximate order reconstruction ($\varepsilon-AOR$).
This research has greatly improved the efficiency of  
reconstruction
and conducted more robust attacks
over previous studies~\cite{kellaris2016generic,grubbs2018pump}.
However, 
to some extent,
they all made some additional assumptions about database.
\Cref{tbl:recrange} presented the key contributions 
and features of reconstruction attacks 
based on the range queries.

\begin{table*}[t]
\tiny
\centering
\begin{threeparttable}
\caption{Range queries: assumptions and complexity of reconstruction attacks}
\label{tbl:recrange}
\setlength{\tabcolsep}{3pt}
\renewcommand{\arraystretch}{1.2}
\newcolumntype{Z}{>{\centering\arraybackslash}m{0.04\textwidth}}   
\newcolumntype{Y}{>{\centering\arraybackslash}m{0.083\textwidth}}  

\begin{tabular*}{\textwidth}{@{\extracolsep{\fill}} Z *{6}{Y} | *{4}{Y}}
\toprule
& \multicolumn{6}{c|}{Assumptions} & \multicolumn{4}{c}{Attack complexity} \\
\cmidrule(lr){2-7}\cmidrule(lr){8-11}
Reference
& Access pattern leakage & Volume leakage & Search pattern leakage
& Rank information leakage & Limited distribution & Dense Database
& FOR\tnote{1} & FDR\tnote{2} & ADR\tnote{3} & Volume Attack\\
\midrule
\cite{kellaris2016generic} & \checkmark & \checkmark &  &  & \checkmark & \checkmark
  & $O(N^{2}\log N)$ & $O(N^{4}\log N)$ &  & $O(N^{4}\log N)$ \\
\cite{grubbs2018pump}      &  & \checkmark &  &  & \checkmark & \checkmark
  &  &  &  & $O(N^{2}\log N)$ \\
\cite{lacharite2018improved} & \checkmark &  &  & \checkmark &  &
  &  & $N\log N + O(N)$ & $O(N)$ &  \\
\cite{gui2019encrypted}    &  & \checkmark &  &  &  &
  &  &  &  &  \\
\cite{grubbs2019learning}  & \checkmark &  &  &  & \checkmark & \checkmark
  & $O(N\log N)$ & $O(N^{2}\log N)$ &  &  \\
\cite{markatou2019full}    & \checkmark &  & \checkmark &  &  &
  & $O(N^{2}\log N)$ & $O(N^{2}\log N)$ &  &  \\
\bottomrule
\end{tabular*}

\begin{tablenotes}
\tiny
\item[1] \emph{Full Ordering Reconstruction} (FOR).
\item[2] \emph{Full Database Reconstruction} (FDR).
\item[3] \emph{Approximate Database Reconstruction} (ADR).
\end{tablenotes}
\end{threeparttable}
\end{table*}


\subsection{Differential Attack}

A key privacy principle is ensuring an individual's participation remains confidential throughout data collection \cite{kifer2011no}, rather than just excluding a single tuple. For instance, deleting an edge in a social network can affect the broader community, inadvertently exposing sensitive information. Motivated by this subtle impact, the following section discusses \emph{differential attacks}~\cite{zhao2020novel}.

\begin{definition}[Differential Attack]\label{def:DA}
Let $D$ be a dataset consisting of $n$ records $D = \{ r_1, r_2, ..., r_n \}$.
$Q$ is a statistical query function.
$Q(D)$ represents the aggregated query result over the dataset.
The goal of the differential attack is to infer information about a specific record $r_t$ 
by comparing query results before and after its removal.
If $P(f(Q(D), Q(D')) = r) > \theta$,
the differential attack is successful,
where $D' = D \setminus \{r_t\}$, 
$P(f(Q(D), Q(D')) = r)$ represents the probability that an attacker correctly infers the presence or attributes of a specific record $r_t$,
and $\theta$ is a predefined confidence threshold.
\end{definition}


\color{black}


With further related research, 
difference attacks are becoming more content-based,
and their applications have also expanded. 
In recent years, 
\textit{Smart Grid} and \textit{Smart Metering} research has attracted increasing attention from both industry and academia~\cite{jia2014humanfactoraware}. 
Protecting user metering data while revealing valuable aggregate statistics 
is crucial in modern smart grid applications. 
Regular meter transmissions to suppliers or centralized databases 
heighten privacy concerns, prompting encryption, anonymization, 
and noise-based obfuscation strategies. 
Nonetheless, privacy disclosure risks persist. 
Studies reveal that individuals' behavior and private information 
can be inferred from electricity consumption records \cite{lam2007novel}. \citet{rottondi2013decisional} proposed a ``decision attack'' 
leveraging temporal correlations in meter readings, 
exposing whether a targeted record is included. 
Later, \citet{jia2014humanfactoraware} introduced Human-factor-aware Differential Aggregation (HDA), 
where an adversary deduces a user's meter readings 
based on presence or absence information.

\section{Attacks on Privacy-preserving Models} 
\label{sec-modelAttack}


Rapid technological advancements have propelled machine learning to new heights, 
garnering widespread attention.
Models for medical diagnosis or facial recognition 
often remain confidential due to sensitive training data or high commercial value. While offering substantial benefits, 
these systems also pose critical privacy risks—such as model inversion \cite{ye2022model}, 
reconstruction \cite{panchendrarajan2021dataset}, 
model extraction \cite{zhu2021hermes}, 
and membership inference \cite{bai2021ganmia}. 
Consequently, there is a growing emphasis on protecting machine learning systems, 
inspiring extensive research into identifying vulnerabilities, 
mitigating attacks, 
and developing robust defenses for privacy-preserving applications.


\subsection{Extraction Attack}

In recent years, 
many private ML models have been deployed with public query interfaces, 
including ML-as-a-Service (MLaaS). 
This setup allows users to train and query private models 
without revealing sensitive information about the models themselves. 
However, balancing public access with confidentiality 
can lead to a heightened risk of ``model extraction'' attacks, 
where adversaries attempt to steal the model's functionality 
using minimal prior knowledge of its training data or parameters.

\color{black}
\begin{definition}[Extraction Attack]\label{def: EA}
The adversary selects a set of queries $ Q = \{ q_1, q_2, ..., q_n \}$ for the ML model $\mathcal{M}$ 
and obtains the corresponding predictions $R = \{ \mathcal{M}(q) \mid q \in Q \} $. 
The adversary then trains a surrogate model $ \hat{\mathcal{M}}$ using the dataset $(Q, R)$
and the goal of extraction attack is that $\hat{M}$ achieves a high similarity with $M$.
\end{definition}

\begin{figure}[ht]
\centering
\includegraphics[width =0.37\linewidth]{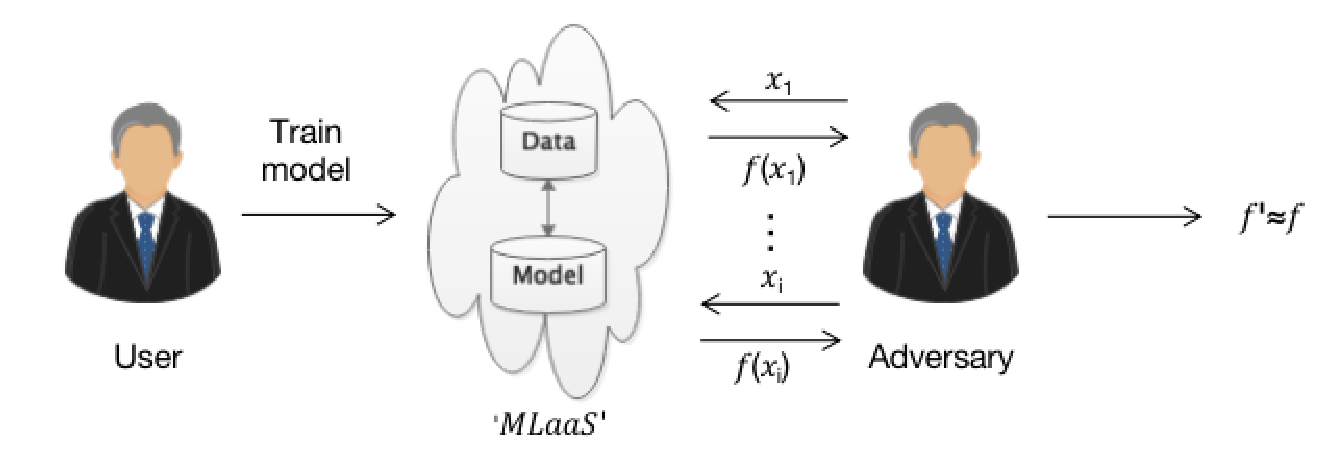}
\caption{The process of ML model extraction attacks: 
The adversary has known part of the model structure or the label information, 
and attempts the unknown parameters via multiple queries to the target. }
\label{fig:extraction}
\end{figure}

\color{black}
In \textit{model extraction attacks},
the adversary attempts to copy or steal secret model parameters 
by sending repeated queries to these prediction APIs,
to compromise model confidentiality
and steal an equivalent or near-equivalent private ML model,
which can be used for subsequent model inversion attacks 
or malicious evasion attacks. 
Furthermore,
due to the huge training costs,
the extraction of sensitive ML models 
may incur huge commercial losses. 
\Cref{fig:extraction} shows the situation 
where the adversary has access to a private black-box model $f$ 
trained by \textit{MLaaS}
and attempts to ``steal'' an approximate model $f'$.

Numerous studies have shown the feasibility of model extraction attacks 
in real-world applications, 
underscoring the risk of private model leakage. 
\citet{tramer2016stealing} introduced equation-solving and decision tree extraction attacks 
exploiting class labels alongside confidence scores returned by ML prediction APIs. 
\citet{wang2018stealing} identified the principle that 
when targeted parameters are retrieved, 
the objective function's gradient approaches zero—revealing an attack 
known as hyper-parameter stealing. 
Minimal changes in the objective function indicate proximity to its minimum, 
exposing critical parameters and structural information.

More recently, \citet{zhu2021hermes} proposed Hermes Attack, 
a large-scale reverse-engineering and semantic reconstruction approach, 
while \citet{luo2022feature} demonstrated that adversaries can leverage Shapley values 
to extract both model parameters and reconstructed model features, 
further emphasizing the severity of this threat.
\textcolor{black}{Recently, 
with the growing popularity of large language models (LLMs), 
a new type of model extraction attack has emerged, 
known as \emph{Model Leeching Attacks}~\cite{birch2023model}. 
The primary goal of this attack is to extract task-specific knowledge from a  
LLM at a low cost 
and use it to build a smaller model with comparable performance.
Model Leeching Attacks achieve this 
by crafting carefully designed prompts to interact with the target LLM, 
collecting its responses, 
and using these outputs as training data for the smaller model. 
Furthermore, 
\citet{more2024towards} has highlighted that in real-world scenarios, 
attackers may employ a range of strategies 
such as training checkpoints and varied prompting techniques
to broaden the attack surface.
This diversified approach can significantly increase the amount of extracted information, 
potentially doubling the original extraction rate, 
and thereby amplifying security and intellectual property 
concerns surrounding LLMs.}


\color{black}
\subsection{Reconstruction Attack}\label{mlrecon}

Unlike the utilizing aggregated dataset query results 
mentioned in \Cref{recAttack}, 
many reconstruction attacks exploit hidden information within ML models 
such as gradients, probability distributions, or activation features 
to infer and reconstruct training data.
The fundamental principle of this type of reconstruction attack is that 
ML models learn and store a significant amount of information 
from training data during the training process. 
Attackers can observe the model's outputs 
or analyze gradient information 
to progressively reverse-engineer and reconstruct the original data.
For example, \citet{panchendrarajan2021dataset} demonstrated how differences in behavior between a generic language model 
and a fine-tuned language model 
can be leveraged to reconstruct the fine-tuning dataset. 
\citet{benkraouda2021image} trained a CNN-based encoder-decoder network 
to convert intermediate representations 
back into the original input images, 
effectively recovering visual data from transformed model outputs.
The comparisons of reconstruction attacks 
for statistically aggregated data 
and privacy-preserving model is shown in \Cref{tbl:recon-target}.

{\setlength{\heavyrulewidth}{0.4pt}%
	\setlength{\lightrulewidth}{0.4pt}%
	\setlength{\cmidrulewidth}{0.4pt}%
	\setlength{\arrayrulewidth}{0.4pt}%
\begin{table} [h] 
\tiny
\arrayrulecolor{black} 
\color{black}
  \centering
  \caption{\textcolor{black}{Comparisons of Reconstruction Attacks on different targets}}
  \label{tbl:recon-target}
  \resizebox{\textwidth}{!}{
  \begin{tabular}{p{60pt}<{\centering}|p{155pt}<{\centering}p{155pt}<{\centering}}
\toprule
     Aspect & Reconstruction Attacks for statistically aggregated data
     & Reconstruction Attacks for privacy-preserving model\\
\midrule
   Attack Target & Query results from databases & ML models \\
   Targeted Data Type & Structured data (e.g., tabular) & Unstructured data (e.g., text, images, audio) and structured data \\
   Attack Mechanism & Infers individual data points by analyzing patterns in aggregated statistics & Extracts sensitive training data by analyzing model outputs, gradients, or internal representations \\
   Effectiveness &  Limited by aggregation mechanisms (e.g., differential privacy can reduce effectiveness) & More effective when models store rich information about the training data \\
   Defense Mechanisms & Differential privacy, data perturbation and query restrictions & Gradient-based protection methods and privacy-enhancing techniques\\
 \bottomrule
\end{tabular}}
\end{table} 
}

Generative Adversarial Networks (GANs) are a common method in reconstruction attacks, 
as attackers can use GANs to generate data that bypasses privacy protection mechanisms. 
Based on GANs, 
\Citet{hitaj2017deep} demonstrated that prototype samples of the training set 
can be generated even under the protection of differential privacy.
For split models, 
which divide computation between edge devices 
and the cloud to protect user privacy, 
\citet{li2023gan} leveraged publicly available data 
with a similar distribution to the target data to train StyleGAN~\cite{karras2020analyzing}. 
By optimizing through his Latent Space Search, 
StyleGAN was able to reconstruct the original data effectively.
Machine Unlearning, 
as a data privacy protection method, 
aims to remove the influence of specific data 
from a trained machine learning model. 
However, 
\citet{zhang2023conditional} proposed a GAN-based reconstruction attack on unlearning, 
demonstrating that even after data has been ``forgotten,'' 
attackers can still reconstruct it. 
By analyzing changes in model parameters before and after unlearning, 
they showed that GANs can effectively exploit these differences 
to recover the supposedly erased data.


FL, 
a federated machine learning paradigm, 
enables collaborative model training across multiple clients 
without direct sharing of private data. 
While FL inherently prioritizes privacy preservation 
as a core design principle, 
emerging research on data reconstruction attacks 
reveals that adversaries can reverse-engineer sensitive training data 
by exploiting exchanged parameters during the FL process~\cite{song2023approximate}.
\citet{yang2022using} introduced a reconstruction attack 
based on highly compressed gradient leakage, 
which can recover training data 
even when the gradient compression rate is as low as 0.1\%.
\citet{song2023approximate} proposed an interpolation-based approximation method 
that enables successful data reconstruction attacks 
in the FedAvg framework.
Additionally, 
\citet{chen2022practical} assumed a scenario 
where the attacker acts as the FL server 
and developed a more efficient gradient matching-based method~\cite{9774939}. 
This approach not only reconstructs the sensitive attributes 
of participants' training data 
but can even infer the sensitive attributes of records 
that are not included in any participant's training dataset.


\color{black}
\subsection{Membership Inference Attack}

When machine learning models are associated with sensitive domains,
such as financial services~\cite{khandani2010consumer} 
or medical research~\cite{backes2016membership},
not only the models themselves,
but the membership information in the training sets
can motivate the privacy risks of individuals.
\color{black}{This type of attack, 
which targets machine learning models by determining 
whether a specific individual record 
is part of a sensitive training dataset, 
is known as a membership inference attack.}

\begin{definition}[Membership Inference Attack]\label{def: MIA}
A membership inference attack targets a specific record $r_t$, 
and the goal of the adversary $A$ is to determine 
whether $r_t$  was used in the training set $D$ by querying the model $M$.
\end{definition}

\color{black}

\begin{figure}[htbp]
\tiny
 \centering
 \subfigure[Black-box]{
  \includegraphics[width=0.4\linewidth]{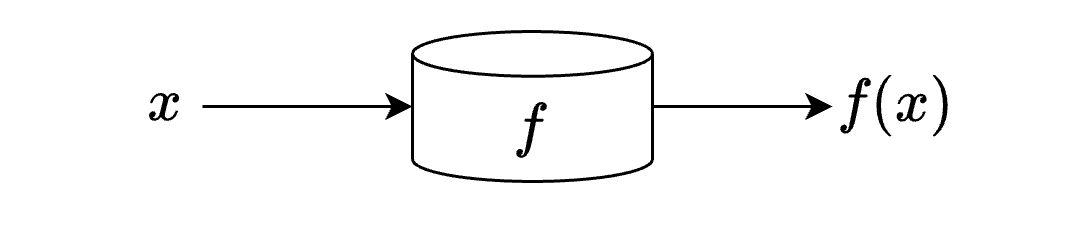}
 }
 \subfigure[White-box]{
  \includegraphics[width=0.5\linewidth]{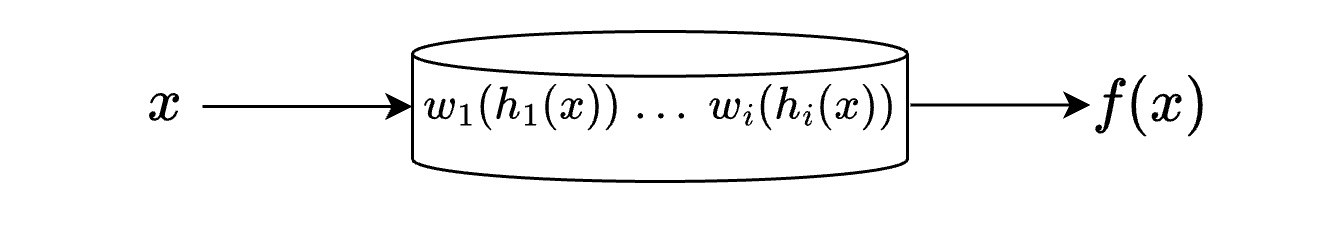}
 }
 \caption{Black-box vs White-box. 
 The former represents a model with unknown model parameters, 
 and the latter represents a model with known internal parameters and structure.}
 
 \label{fig:blackwhite} 
\end{figure}

As we know,
in the process of constructing machine learning models,
a lot of sensitive data,
including individuals' transactions and preferences,
medical health records, or face images,
are used as training data.
There is some possibility that
black-box models or white-box models
unintentionally reveal the secrets of private training data 
either by the specific predictive behavior
or the details of their structures and parameters.
In machine learning,
the black-box setting is referred to as the condition 
where the adversary can only obtain the model's output results under given inputs,
while the white-box setting represents the condition 
under which the adversary can learn almost all the secret parameters 
and internal structure of the model,
and both are common in practical application scenarios. 
\Cref{fig:blackwhite} shows the visual difference between 
the two settings above.

Many privacy researchers assert that privacy-preserving ML models risk 
revealing training-set membership in both black-box and white-box scenarios. 
\citet{shokri2017membership} first demonstrated membership inference attacks 
against ML-as-a-Service (MLaaS) via a novel shadow training technique, 
training an ``attack model'' to infer membership solely from target-model outputs. 
However, 
they assumed identical training data 
and model structure for both the target and shadow models, 
limiting the attack scope.

\citet{liu2019socinf} proposed \textit{SocInf}, 
a more effective membership inference method that leverages neural networks 
and GANs to detect prediction discrepancies. 
By repeatedly querying target models to rebuild posterior vectors, 
adversaries can achieve high attack performance under black-box conditions. 
Addressing data scarcity for training the attack model, 
\citet{bai2021ganmia} presented \textit{GANMIA}, 
a framework that uses GANs to generate synthetic samples, 
improving both robustness and accuracy of black-box membership inference attacks.

All these works underscore the feasibility of membership inference attacks 
against ML models in black-box environments, 
where adversaries only access model predictions.
\textcolor{black}{
While black-box attacks remain practical, 
they can be more challenging against deep learning models 
with large parameter sets 
due to their improved generalization ability 
and limited output information~\cite{duan2024membership}.
However, 
recent work has demonstrated 
that optimized black-box attacks can still be successful against 
large-scale models by leveraging specific inference techniques.
For example, 
MIA on large-scale multi-modal models leverages cosine similarity
and augmentation-enhanced techniques to successfully infer membership~\cite{ko2023practical}. 
In in-context learning, 
black-box MIA attacks rely solely on the model's generated text 
to infer membership status. 
This can be achieved through the brainwash attack, 
which involves feeding the model incorrect information 
and observing whether it ``insists on the correct answer,'' 
as well as the hybrid attack, 
which combines the brainwash attack with the repeat attack 
for enhanced effectiveness~\cite{wen2024membership}.}

\Citet{sablayrolles2019white} proposed a Bayes optimal strategy
for membership inference   
with mild assumptions
on the distribution of the parameters,
that depended only on the loss function rather than the parameters.
They also showed that the state-of-the-art 
MIA methods
can closely replace the Bayes optimal strategy~\cite{shokri2017membership,yeom2018privacy}. 
Moreover, 
in recent studies,
MIAs are successful applied to various models~\cite{duan2023diffusion,ko2023practical}.
\citet{duan2023diffusion} first demonstrated 
that diffusion-based graph generation models are also vulnerable to MIA.
\citet{ko2023practical} conducted MIA experiments on large-scale multimodal models 
and demonstrated that even advanced multimodal models like CLIP 
reveal significant privacy issues.
The comparison of MIA in different settings 
is presented in ~\Cref{tbl:bwcom}.

\begin{table} [h] 
\tiny
\color{black}
  \centering
  \caption{\textcolor{black}{MIA under two settings}}
  \label{tbl:bwcom}
  \resizebox{\textwidth}{!}{
  \begin{tabular}{p{50pt}<{\centering}|p{150pt}<{\centering}p{165pt}<{\centering}}
\toprule
     Aspect & Black-Box MIA
     & White-Box MIA\\
\midrule
   Adversary Access & Only input-output access (e.g., prediction probabilities, confidence scores) & Full access to model parameters, gradients, and internal activations \\
   Attack Methodology & Statistical inference or training shadow models to mimic target behavior & Exploiting gradient information or analyzing parameter sensitivity to specific samples \\
   Feasibility & More practical, applicable to commercial APIs & Requires access to model internals, more restrictive in real-world scenarios \\
   Effectiveness & Moderate; effectiveness may be reduced in models with improved generalization capabilities & Higher\\
   Attack Cost & Lower computational cost & Higher computational cost (gradient analysis and other techniques)\\
 \bottomrule
\end{tabular}}
\end{table} 



\subsection{Model Inversion Attack}

It is generally believed that 
\textit{membership inference} 
and \textit{attribute inference} are somewhat related. 
\citet{yeom2018privacy} discussed the issue and
further supported the relationships among privacy risks
of ML models,
over-fitting,
and influence,
including their impact on membership inference and attribute inference.
Overfitting stands as a powerful enabler for attackers, 
facilitating not only membership inference but also, 
under certain conditions, 
the execution of attribute inference attacks. 
Its influence extends to both areas, 
underscoring the critical need for vigilance and robust defense mechanisms.
In the attribute inference attack targeted at ML models,
the models and incomplete information of some records
are exploited to infer the missing sensitive attribute of 
the targeted record. 
Specifically,
the adversary aims to infer the exact value 
of the hidden sensitive attributes of specifically targeted records
by analyzing the released data together with models,
which is considered successful 
if the inference is correct with sufficient probability.
A major research topic related to 
attribute inference attack against ML models 
was \textit{model inversion attacks},
where the adversary leverages the predictions of ML models
to infer sensitive attributes used as input~\cite{dibbo2023sok}.

 \color{black}
\begin{definition}[Model Inversion Attack]\label{def: MInA}
The adversary selects a set of queries $ Q = \{ q_1, q_2, ..., q_n \}$ for the ML model $\mathcal{M}$ 
and obtains the corresponding predictions $R = \{ \mathcal{M}(q) \mid q \in Q \}$.
The goal of a model inversion attack is to 
infer information about the training data 
$X$ 
by reconstructing inputs or sensitive attributes 
that are consistent with the model’s outputs.
\end{definition}


\begin{table}[t]
\Large
\centering
\caption{The comparison of four model attacks}
\label{tbl:threemodelattack}
\resizebox{\textwidth}{!}{
\begin{tabular}{m{150pt}<{\centering} m{120pt}<{\centering} m{110pt}<{\centering} m{120pt}<{\centering} m{110pt}<{\centering} m{140pt}<{\centering}}
\toprule
References &  Attack Type & Motivation & Attack target & Auxiliary information & Countermeasures \\
\midrule
\citet{tramer2016stealing}      & Model extraction      & Model theft              & Model parameters/structure & $\times$     & DP \\
\citet{fredrikson2015model}     & Model inversion       & Attribute inference      & Unknown sensitive attributes & \checkmark & DP / Knowledge Distillation / Regularization \\
\citet{benkraouda2021image}     & Reconstruction        & Record reconstruction    & Input data                 & \checkmark & DP \\
\citet{shokri2017membership}    & Membership inference  & Participation disclosure & Target training data       & Both       & Knowledge Distillation / Regularization \\
\bottomrule
\end{tabular}}
\end{table}

\color{black}


 \color{black}
The initial research on model inversion attacks exploiting 
\emph{Maximum A Posterior} (MAP) estimators
was launched by~\citet{fredrikson2014privacy}.
In this study,
the adversary was given incomplete auxiliary information 
about a patient's historical medical data
and the main purpose was to 
accurately infer the genotype of the targeted patient  
through ML predictive model trained 
on similar medical history datasets.
They intended to find the correlations among the attack target,
the model output and other attributes,
while providing a general inversion algorithm.
However,
their attack was limited by the fact that 
the inferred sensitive features
could only be derived from a small domain. 
To demonstrate the broader risk of inversion attacks,
\citet{fredrikson2015model} presented 
the novel white-box inversion attack
exploiting the confidence information exposed by MLaaS APIs,
which represents the likelihood
from the intermediate layer of the model.

For FL, 
\citet{zhu2019deep} showed how  we can obtain the training data 
by developing a deep leakage from the gradients.
\citet{luo2022effective} presented another idea of gradient inversion 
by reconstructing training data. Using transfer learning, 
\citet{ye2022model} proved that the inversion attacks fall apart
when targeting the student model. 
However, 
this will recover the training data successfully 
when targeting the teacher model.
Moreover, 
\textcolor{black}{
\citet{qi2023model} proposes 
\emph{Dynamic Memory Model Inversion Attack} (DMMIA) (a variant of MIA)
that leverage knowledge gained during training 
to induce the model to generate diverse new samples,
which achieved highly effective attack performance.
Its ability to generate diverse samples from learned knowledge 
makes it a powerful attack technique, 
emphasizing the need for stronger privacy-preserving mechanisms 
such as differential privacy and knowledge distillation.
}
\textcolor{black}{
The above-mentioned model inversion attacks 
primarily focus on black-box and white-box settings. 
However, 
\citet{nguyen2023label} demonstrated that even in scenarios 
where only labels are available, 
successful attacks can still be carried out through knowledge transfer. 
This finding highlights that limiting API outputs to hard labels 
does not fully eliminate privacy leakage risks. 
}
\Cref{tbl:threemodelattack} summarized classic model attacks 
that researchers are most concerned about.

\section{Countermeasures and Enablers}\label{sec-counter}

In this section, we review existing countermeasures to the aforementioned attacks and highlight potential enablers of resilience. These attacks pose not only technical challenges but also serious risks for organizations, including regulatory scrutiny and litigation over inadequate data protection. Importantly, privacy-preserving mechanisms are not one-size-fits-all; their effectiveness depends on the nature of the data and the threat model. Measures against different categories of attacks are summarized in \Cref{preservean}, \Cref{preservesta}, and \Cref{securepri}. \Cref{ppr} provides a brief overview of data protection regulations motivating these techniques, while \Cref{tbl:countermeasure} maps major attack types to corresponding countermeasures.
\color{black}

\begin{table}[t] 
\arrayrulecolor{black} 
\color{black}
  \centering
  \caption{\textcolor{black}{Attacks vs Countermeasures}}
  \label{tbl:countermeasure}
\resizebox{\textwidth}{!}{
\centering
\begin{tabular}{c|ccc} 
\midrule
Attack Target                              & Attack Method               & Possible Countermeasures   & {Data Type with References}  \\ 
\midrule
\multirow{2}{*}{Anonymous Data}            & Linkage Attack            & \begin{tabular}[c]{@{}c@{}}Perturbation (Generalization/Bucketization/Suppression)\\Randomization (Differential privacy)\\Partitioning and Clustering \end{tabular} &  
\begin{tabular}[c]{@{}c@{}}Trajectory data~\cite{ghasemikomishani2016pptd}\\Tabular data~\cite{chamikara2020efficient}\\
Graph Data~\cite{wang2022safeguarding}\\
 Dynamic data~\cite{anjum2017tau}
\end{tabular}                                  \\
\cline{2-4}
& Structural Attack           & \begin{tabular}[c]{@{}c@{}}
Perturbation (modifying vertices and edges)\\
Randomization (Differential privacy)
\\Partitioning and Clustering\end{tabular}                                                      &  \begin{tabular}[c]{@{}c@{}}Graph data~\cite{xue2012delineating}\\Graph data~\cite{ji2016graph}\\Graph data~\cite{jiang2015novel} \end{tabular}                                                \\ 
\midrule
\multirow{3}{*}{Statistical Data}          & Reconstruction Attack~      & \begin{tabular}[c]{@{}c@{}}Query Auditing \\ Modifying Query Processing \\ Adding Noise (Output noise injection)\end{tabular}                                 
&   \begin{tabular}[c]{@{}c@{}} Tabular data~\cite{khan2020quantitatively}
\\ Trajectory data~\cite{fanaeepour2015case}  \\Graph data~\cite{guan2023efficient} \end{tabular}                                                     \\
\cline{2-4}
& Differential Attack         & Disclosure Avoidance~                          &    Tabular data~\cite{vaidya2013identifying}       \\
\midrule
\multirow{4}{*}{Privacy-preserving Models} & Extraction Attack           & Differential Privacy                                                                                    &      Model parameters~\cite{yan2021monitoring}   
 \\
 \cline{2-4}
& Reconstruction Attack  & Differential Privacy & Model parameters~\cite{kaissis2023bounding}
\\
\cline{2-4}
& Membership Inference Attack & \begin{tabular}[c]{@{}c@{}}Regularization\\Knowledge Distillation\end{tabular}                                                    & \begin{tabular}[c]{@{}c@{}}Model parameters~\cite{li2021membership} \\ Model parameters~\cite{zheng2021resisting}  \end{tabular}                                                \\
\cline{2-4}
& Inversion Attack            & \begin{tabular}[c]{@{}c@{}} Differential Privacy\\ Regularization\\Knowledge Distillation\end{tabular}                                                                          &    
\begin{tabular}[c]{@{}c@{}} Model parameters~\cite{zhang2020broadening}\\ Model parameters~\cite{wang2021improving}\\ Model parameters~\cite{chen2025kdrsfl}  \end{tabular}                                              \\
\midrule
\end{tabular}}
\end{table}

\subsection{Preserving Anonymous Data}\label{preservean}

In addition to minimizing the hidden risks of privacy disclosure, 
implementing higher standards for re/de-identification may result in 
research data being de-identified, 
thus decreasing its utility. 
The deviation of the analytical research results 
using de-identified databases on the \textit{edX} platform
has been proved in~\citet{daries2014privacy}.
Because of the tension between open data and privacy,
compared to anonymous datasets,
better solutions are urgently needed.
Ongoing research works have identified many best practices
to prevent re-identification; 
see~\cite{jiang2015novel,
wang2022safeguarding, wang2023pseudo} 
and the references therein.

\subsubsection{Partitioning and Clustering}

To reduce the possibility of using quasi-identifiers 
to match particular individuals,
partition-based methods focus on dividing the database
into different disjoint groups that fulfil certain criteria,
such as K-anonymity and its variants. 
K-anonymity requires each record to be indistinguishable,
because at least $k$ records have the same attributes~\cite{sweeney2002kanonymity}. 
Its variants include \texttt{l-diversity}~\cite{machanavajjhala2007diversity}, 
\texttt{t-closeness}~\cite{li2006t},
\textcolor{black}{\texttt{$\tau$-safety}~\cite{anjum2017tau}}, 
etc, 
with the idea of data desensitization. 
Although these schemes can help protect against re-identification,
it has been found that 
they are vulnerable to
\textit{composition attacks}~\cite{ganta2008composition}
because of the precise information released.
Following the same principle,
to protect the private information of individuals in the graph data,
clustering-based methods attempt to cluster vertices and 
edges in the graph into different groups,
as well as anonymizing the subgraphs into super-vertices~\cite{jiang2015novel}. 
\textcolor{black}{
For knowledge graphs, 
based on the concept of K-anonymity,
\citet{hoang2023protecting} introduced the notion of anonymization distance, 
which comprehensively considers attribute similarity between users, 
structural similarity of relationships, 
and differences in the k-value (the level of anonymization). 
By integrating clustering and graph generalization techniques, 
they proposed the personalized cluster-based knowledge graph anonymization method. 
This approach not only ensures privacy protection 
but also generates high-quality clusters, 
thereby maintaining data utility while safeguarding sensitive information.
}


\subsubsection{Randomization}

The success of re-identification depends on
how unique the quasi-identifiers are.
So randomly introducing uncertainty into the raw data
or quasi-identifiers,
such as \textit{randomized response}
and \textit{local DP}~\cite{dwork2013algorithmic},
can prevent de-anonymization attacks.
\textcolor{black}{The key idea of local DP is to 
introduce controlled noise into the data before sending it to the server, 
ensuring that individual data points remain private 
while still allowing useful aggregate analysis.}
For example,
\citet{wang2022safeguarding} demonstrated how to use local DP 
with low noise for online social networks.

\subsubsection{Utility and privacy analysis}
Partitioning, clustering, and randomization techniques impact data utility in distinct ways, and their effects can be rigorously quantified using specific metrics. Partitioning and clustering protect privacy by generalizing or aggregating quasi-identifiers, but this can reduce data discriminability. In graph-structured data, subgraph clustering may distort original relationship patterns~\cite{chang2016privacy}, degrading the accuracy of tasks such as network analysis, link prediction, and recommendation systems. Relevant utility metrics for these methods include information loss\cite{hoang2023protecting}, query accuracy\cite{wang2014clustering}, and statistical similarity~\cite{kim2024privacy}.

Randomization methods protect privacy by injecting controlled noise into the original data, reducing the risk of re-identification. While effective—particularly under tight privacy budgets—this approach often compromises analytical accuracy. In aggregate analyses (e.g., means, histograms, frequency counts), noise introduced under LDP can cause substantial deviations from true statistics~\cite{yang2020local}. These deviations are especially pronounced when the number of aggregated records is small, as the relative impact of noise increases, limiting the reliability of statistical modeling and inference.

\color{black}

\subsection{Preserving Statistical Data}\label{preservesta}

Attacks on statistical aggregate data 
summarized earlier show that 
combining a certain number of specific private query results 
with database-related auxiliary knowledge 
can bring the risks of privacy leakage to a certain extent.
Here,
we discuss some generic kinds of countermeasures 
that help to reduce  the possibility of 
conducting successful attacks,
thus improving the resilience of statistical aggregate publishing 
against these attacks.

\subsubsection{Disclosure Avoidance}

Intuitively,
releasing less statistical data or carefully processed data
is a reasonable countermeasure against reconstruction attacks.
For example,
\textit{bucketization},
\textit{generalization} and \textit{cell suppression}
~\cite{xu2014survey}.
However,
just as discussed above,
these methods with certain limitations 
fail to protect privacy against powerful adversaries effectively 
~\cite{garfinkel2019understanding}.

\subsubsection{Adding Noise}

Curators can choose to add noise to the original data or to the published statistical data,
which are called \textit{input noise injection} 
and \textit{output noise injection} respectively.
Similarly,
dummy records can be added to databases.
But we need to keep a balance between data utility and privacy 
while perturbing the databases,
as discussed above.
More formal privacy models,
such as \textit{DP}~\cite{dwork2013algorithmic},
can add noise on the data or dummy records.
A large number of research works have shown that 
\textit{DP} is a more principled way 
to achieve strict and provable privacy guarantees in released statistics 
by adding random noise into the exact statistical values. 


\subsubsection{Query Auditing}

In an attempt to achieve effective privacy-preserving,
privacy statistical queries issued by the adversary 
should be continuously checked  to prevent information leakage.
Given the privacy definition and the answers to past queries,
any new query that leads to private information disclosure 
should be denied.
\citet{khan2020quantitatively} introduced a quantitative privacy metric called \textit{PriDe}, 
which calculates the privacy risk score when querying a private database.
This metric can be deployed in an interactive query environment 
to monitor and protect data privacy.
Additionally,
\textit{SAIL DATABANK}, 
a popular customizable analysis platform, 
employs a privacy governance model to provide anonymous data across various fields, which can avoid privacy disclosure~\cite{jones2019profile}. 
But we should know that 
query denials themselves can also 
leak privacy,
and sometimes, the denials will also reduce the data utility.
As a result,
more efficient \textit{non-deniable auditing} 
has been proposed~\cite{lu2009efficient},
which serves as a starting point for privacy research 
and provides an important reference for future related research.

\subsubsection{Modifying Query Processing}

Certain restrictions on private queries may also 
reduce the effectiveness of reconstruction attacks.
For example,
if the server batches some queries together 
instead of processing them individually,
it can prevent the volume of each query from being revealed,
thus preventing volume attacks from succeeding~\cite{grubbs2018pump}.
Another viable practice is to set a lower bound on the width of range queries,
which can also prevent the reconstruction of exact counts of individual records. 
Query algorithms with privacy-preserving features 
are also promising research directions.
\citet{li2020adaptively} proposed 
a privacy-preserving dynamic conjunctive query framework 
that satisfies adaptive security, 
scalable index size, 
and efficient query processing at the same time. 

\subsubsection{Utility and privacy analysis}

Disclosure avoidance techniques generally offer limited privacy protection, 
as they are designed to retain high utility~\cite{garfinkel2019understanding}. 
Methods such as query auditing and modified query processing 
can mitigate query-volume-based attacks 
but may reduce the granularity and flexibility of query results. 
Common utility metrics in these approaches include 
query accuracy~\cite{grubbs2018pump} 
and query efficiency~\cite{lopuhaa-zwakenberg2021privacyutility}.
DP provides strong and mathematically provable privacy guarantees.
It has two key parameters, privacy budget ($\epsilon$) 
and the probability of a privacy failure ($\delta$)~\cite{zhu2017differentially}.
Smaller $\epsilon$ implies stronger privacy 
but higher noise and lower data utility.
$\delta$ allows for a small chance 
that privacy may not be strictly preserved.
When privacy requirements are high (smaller $\epsilon$), 
data accuracy tends to degrade significantly 
due to the increased noise~\cite{lan2022distributed}.

\color{black}
\subsection{Securing Privacy-preserving Models}\label{securepri}

The attacks discussed above 
fully illustrate that simply hiding parameters 
or model structures is far from sufficient 
to fully achieve robust privacy-preserving of models.
In the following,
we conclude common approaches 
to defensive strategies 
against various privacy attacks,
which can guide future research 
towards more efficient and robust privacy-preserving models.

\subsubsection{Perturbation of Gradients}
The inaccuracy of gradients can hinder adversaries from inferring true attributes, motivating several gradient perturbation methods~\cite{yin2021defending,yan2021monitoring}. For instance, \citet{yin2021defending} replaces updated gradients with their average values, which, together with data balancing, helps resist attacks from diverse adversaries. A widely adopted approach is DP, which injects calibrated noise into data or model parameters to limit privacy leakage while preserving query accuracy. DP ensures that slight changes in model outputs do not reveal whether a specific record was used in training, thereby mitigating MIA and reducing the risk of model extraction or inversion attacks.

Evaluation methods have been proposed to quantify DP’s effectiveness: \citet{park2019attackbased} introduced an attack-based framework to assess resistance against model inversion attacks, providing a benchmark for privacy-preserving deep learning. Similarly, \citet{huang2024gradient} show that gradient perturbation, such as DP noise, can partially defend against image reconstruction attacks using conditional diffusion models, although its effectiveness depends heavily on the magnitude of the noise. Gradient perturbation can also counter query flooding-based model extraction; \citet{yan2021monitoring} proposed a monitoring-based DP mechanism that dynamically adjusts noise according to query behavior, enhancing model protection.
\color{black}

\subsubsection{Reduce Overfitting}

The success of most privacy attacks targeted 
at privacy-preserving ML models is largely due to 
the inevitable overfitting characteristics inherent in the ML models,
which will increase the prediction difference 
between data that the targeted models have never ``met'' before 
and the data on which the models have been trained.
It's believed that exploiting this difference can greatly 
increase the risk of privacy violations.
\citet{mukherjee2020protecting} developed 
an effective GAN model privacy-preserving architecture called \textit{privGAN}, 
which proved that preventing overfitting can largely 
prevent membership inference attacks.
\Citet{li2021membership} found that 
the attack accuracy was closely related to 
the model generalization gap 
and suggested that 
by appropriately reducing the training accuracy and
combined with the confusion training method 
the model's defense against membership inference can be strengthened. 
As a result,
using \textbf{regularization} techniques
to avoid overfitting will effectively defend against such attacks~\cite{mukherjee2020protecting,li2021membership,yeom2018privacy}.

\subsubsection{Knowledge Distillation}

Knowledge distillation serves as a promising privacy-enhancing mechanism, 
enabling the deployment of privacy-aware ML models 
with improved security against data leakage and adversarial threats
In knowledge distillation, 
a large, complex teacher model transfers its learned knowledge to a smaller, 
more lightweight student model, 
typically by distilling soft labels or probability distributions instead of exposing raw training data~\cite{kundu2021analyzing}. 
This indirect knowledge transfer significantly reduces the risk of membership inference attacks and model inversion attacks, 
as the student model does not have direct access to the original training dataset.

\textit{Private Aggregation of Teacher Ensembles} (PATE) is a classic method 
that leverages knowledge distillation to enhance privacy protection~\cite{papernot2016semi}. 
It employs an ensemble of teacher models trained on disjoint data subsets, 
which vote on labels. 
DP mechanism obscures individual contributions 
before the student model distills knowledge from the noisy aggregated outputs, 
ensuring a differentially private approximation of the original dataset.
Besides,
privacy-preserving knowledge distillation can be particularly useful 
in settings like FL.
To defend against malicious client attacks, 
\citet{park2023feddefender} proposed a method to extract valid knowledge 
from a potentially poisoned global model while filtering out incorrect information, 
thereby enhancing the robustness of FL.
To mitigate model inversion attacks, 
\citet{chen2025kdrsfl} combined adversarial training with knowledge distillation, 
enabling the student (client) to mimic the teacher model's outputs. 
This approach effectively reduces the success rate of model inversion attacks 
while maintaining high model accuracy.




\subsubsection{Utility and privacy analysis}

Different privacy-preserving methods intervene in model training in distinct ways, 
leading to varying impacts on utility depending on the underlying mechanisms. 
Regularization can reduce overfitting 
and lower the success rate of privacy attacks while maintaining good utility. 
But excessive regularization may 
result in underfitting and degraded performance.
Various noise-injection strategies are suited to 
different threat models and application settings.
Adding noise to gradients enables privacy: 
server-side perturbation (DP-SGD) offers a stronger utility trade-off 
under a trusted trainer, 
whereas client-side perturbation (LDP) protects against an untrusted aggregator 
but usually harms utility more~\cite{kato2022olive}.
For PATE,
multiple teacher models trained on disjoint data partitions produce predictions 
that are noisily aggregated, 
and the student learns from these aggregated outputs 
without direct access to raw training data. 
Its utility is influenced primarily by the aggregation noise, 
the number and quality of teachers, the query budget, 
and student-side choices (capacity, distillation temperature)~\cite{papernot2016semi}.
For model-level privacy, 
evaluation should pair standard utility metrics 
with privacy accounting and empirical attack metrics~\cite{yin2021defending}, 
to make the privacy–utility trade-off explicit.

\subsection{Privacy Protection Regulations}\label{ppr}

Privacy-preserving techniques are not only motivated by technical concerns 
but are increasingly driven by legal and regulatory requirements. 
Data protection regulations such as the \emph{General Data Protection Regulation} (GDPR)~\footnote{https://eur-lex.europa.eu/legal-content/EN/TXT/?uri=CELEX:32016R0679} in the EU, 
the \emph{California Consumer Privacy Act} (CCPA)~\footnote{https://oag.ca.gov/privacy/ccpa} in the US, 
\emph{Personal Information Protection Law of the People’s Republic of China} (PIPL)~\footnote{https://personalinformationprotectionlaw.com/}, 
and \emph{Act on the Protection of Personal Information}~\footnote{https://www.japaneselawtranslation.go.jp/en/laws/view/4241/en} in Japan, 
impose strict obligations on how personal data is collected, processed, and protected.

GDPR emphasizes principles such as data minimization, 
purpose limitation, privacy by design and by default, 
and accountability. 
These principles align closely with 
the use of anonymization techniques, differential privacy, 
and model-level protection, 
which reduce the exposure of raw data 
and limit the identifiability of individuals.
In particular, 
the GDPR’s \textit{Recital 26} distinguishes 
between personal data and anonymous data, 
clarifying that data anonymized in a way that individuals are 
no longer identifiable is no longer subject to the regulation. 
This incentivizes organizations to implement strong anonymization 
and privacy-preserving mechanisms not only to mitigate risk, 
but also to reduce regulatory obligations.
Furthermore, 
compliance with these regulations often requires organizations 
to demonstrate that appropriate technical 
and organizational measures have been taken. 
The techniques discussed in \Cref{sec-counter} 
provide a foundation for such measures, 
supporting both regulatory compliance and resilience against privacy attacks.

\color{black}
\section{Future Research Directions} \label{sec-reandch}

Although some potential privacy risks have been identified,
this section is devoted to the emerging issues  
related to other privacy attacks 
that require further research attention.

\subsection{Privacy Attack on Differential Privacy}

\emph{Differential Privacy} (DP) is the most successful
privacy-preserving mathematical framework 
due to its lightweight and easy implementation without prior knowledge.
Recent research~\cite{zhu2017differentially, tramer2016stealing}
on DP open a feasible way
to achieve strong and provable privacy guarantees.
DP limits what can be learned from aggregated query results 
about privacy statistics databases,
and reduces the possibility of privacy violations
by ensuring that 
the presence of any record in the database  
has a statistically negligible effect.
Nevertheless,
due to its complicated nature of strict privacy standards
and its assumption of complete trust in the data curator,
the general DP 
has been shown to be inadequate against 
some specific privacy attacks~\cite{yang2023model}.
\citet{yang2023model} proposed \emph{Model Shuffle Attack},
which strategically shuffles and scales model parameters 
to perform stealthy model poisoning in DP-enabled FL.



Addressing the issue of requiring a trusted data curator in DP,
\emph{Local DP} (LDP) has been applied to many large distributed systems in various fields
to collect and analyze sensitive user data,
such as Google's \textit{RAPPOR}, 
\textit{Prochlo} and Apple's \textit{iOS}.
\textcolor{black}{\citet{cheu2019manipulation} emphasized a fundamental limitation
that LDP is highly vulnerable to manipulation attacks.}
The first systematic research work into \textit{data poisoning attacks} against LDP
for heavy-hitter, 
identification was done in~\cite{cao2021data}.
\textcolor{black}{Therefore, 
future research should focus on exploring novel privacy attacks,
particularly innovative attacks targeting DP, LDP and their variants~\cite{song2025towards,mironov2017renyi}. 
Such efforts will drive the continuous advancement of DP 
in both privacy protection and security defense.}



\subsection{Privacy Attack on Spatio-temporal Data}

With the development of location-aware technology 
and the outbreak of the COVID-19 epidemic, 
people's trajectories and whereabouts are being collected and 
released for various reasons. 
Besides,
many location-based applications and services have continued to
collect and share human mobility traces for traffic forecasting, 
urban road planning,
or provding other real-time life-enriching experiences
~\cite{ma2019trafficpredict}. 
If these trajectory data are revealed, 
it will pose a major threat to privacy, 
such as revealing travel records or
sensitive locations visited.
Therefore, 
privacy of location-based mobility traces is becoming 
increasingly popular and of great concern.
Currently,
popular countermeasures include 
Release of trajectory data based on virtual trajectory and DP.

The vast majority of previous research on location privacy
has used traditional point-based location perturbations that
obfuscate or perturb each location point 
by using a fake location or a cloaked region, 
which is vulnerable to inference attacks
and has difficulty ensuring space utility.
\citet{gursoy2018utilityaware} presented a Bayesian inference attack,
partial sniffing attack and outlier linkage attack on trajectory data.
The privacy risks and challenges of \textit{space crowdsourcing},
a popular platform for collecting and disseminating spatio-temporal information,
have been highlighted by~\citet{tahmasebian2020crowdsourcing}. 

\subsection{Privacy Attack over Existing ML}

Various emerging machine learning techniques 
such as multi-view learning and federated learning~\cite{pokhrel2023modeling}, 
aim to improve the efficiency or privacy of the training process. 
However, despite the search for optimised models, 
significant privacy risks remain, 
especially in multi-view learning, which 
despite its popularity in the era of big data, 
has not been fully explored for privacy protection. 
\citet{xian2020multiview} presented a framework 
for de-anonymising network data based on \emph{Multi-View Low-Rank Coding} (MVLRC), 
highlighting the inefficiency of traditional privacy-preserving techniques 
when applied to multiview data.
\citet{mai2023privacy} demonstrated that 
even with a small amounts of DP noise on multi-view learning gradients,
individual information can be reconstructed,
hence they provide a novel \textit{PrivMVMF} framework based on Homomorphic encryption.


FL 
is also considered to be the latest breakthrough 
in the privacy-preserving machine learning research,
in which models are trained in a decentralized manner 
by independent data curators,
preventing their private data from being shared with to others.
However,
many existing FL frameworks have been shown to be
vulnerable to privacy risks,
and it may not always provide sufficient privacy guarantees.
The attack by malicious servers against the \textit{FL}
has been discussed by~\cite{wang2018inferring},
which explored the user-level privacy leakage for the first time.



\subsection{Privacy Attack over Machine Unlearning}
\textcolor{black}{
ML models sometimes unintentionally memorize sensitive training data, 
which can lead to privacy breaches. 
To address this issue, 
\emph{machine unlearning} has been proposed~\cite{10.1145/3603620}. 
It allows models to selectively forget specific knowledge 
or data points without requiring full retraining, 
thereby removing sensitive information while preserving model performance.} 
Many studies have demonstrated that machine learning can remove the
features of sensitive training data successfully without the need to retrain,
while maintaining model performance~\cite{wu2022puma,yan2022arcane}. 
\citet{wu2022puma} proposed \emph{Performance Unchanged Model Augmentation} (PUMA),
which can analyze how individual data influences a model's overall accuracy and behavior,
and if a labeled data point were to be removed, 
it seeks to mitigate negative impact 
through appropriately reweighing the the remaining data.
\citet{yan2022arcane} realized \emph{Exact Machine Unlearning} 
by converting the learning objective 
into multiple one-class classification tasks 
and independently training sub-models on each partitioned sub-dataset.
This method can efficiently remove data that needed to be forgotten and reduce the cost of retraining.


\subsection{Towards Detecting Privacy Violation}

Despite extensive efforts within the privacy research community 
to develop robust methods for detecting privacy violations, 
effectively identifying latent leaks remains challenging. 
Researchers have proposed diverse approaches 
to detect various privacy risks~\cite{dou2012detecting, ding2018detecting, juuti2019prada}. 
\citet{dou2012detecting} specifically addressed privacy violations in multi-view publishing, 
examining breaches of traditional \textit{k-anonymity}. 
Subsequent advancements in DP 
have introduced sophisticated privacy-preserving algorithms, 
though many purportedly DP-compliant solutions 
contain vulnerabilities undermining their privacy guarantees. 
Consequently, 
rigorous evaluation and detection mechanisms 
are urgently required to validate DP effectiveness. 
\citet{ding2018detecting} developed a counterexample generator 
leveraging classical statistical tests to effectively identify DP violations. Meanwhile, 
\citet{juuti2019prada} introduced \textit{PRADA}, 
a robust detection framework for deep neural network model extraction attacks, 
which analyzes deviations in prediction query distributions. 
Additionally, measuring privacy leakage risks remains critical; 
\citet{wagner2018technical} provided a comprehensive survey 
on privacy metrics from various perspectives, 
serving as an essential reference framework for evaluating privacy levels.


\section{Conclusion} \label{sec-conclusions}

The vast personal data collected and shared by institutions has become a valuable resource for knowledge discovery and data analysis, contributing significantly to improvements in living standards. However, such data sharing also enlarges the attack surface, increasing the risk of privacy breaches. Protecting the confidentiality of personal information has therefore become an increasingly critical challenge.

We conducted a comprehensive review of recent literature on privacy attacks targeting anonymous data, statistical aggregates, and privacy-preserving models. Our analysis systematically categorized attack methodologies, evaluated the robustness of existing defenses, and highlighted emerging strategies. Additionally, we reviewed countermeasures to improve understanding of privacy mechanisms and identify research gaps. A key challenge remains the trade-off between data utility and privacy protection, particularly in large-scale data applications. Our assessment emphasizes resilience to privacy attacks as a central research priority, guiding future work to address emerging threats while optimizing privacy-utility balance.

Overall, 
considering all kinds of possible attacks,
the protection of privacy remains a major challenge.
Given the broad spectrum of privacy concepts,
the survey is inevitably limited in its context and scope,
but we believe that 
such study will add value and develop insights into 
different resilience issues against privacy attacks,
and it will also promote the exploration of novel and robust solutions 
to the privacy-preserving of sensitive information.